\documentclass[a4paper,11pt]{article}
\usepackage{jcappub} 

\newif\ifdily\dilytrue  

\usepackage{cleveref}
\usepackage{amsmath}
\usepackage{tabularx}
\usepackage{booktabs}
\usepackage{subcaption}

\newcommand{\Prob}{\text{P}}           
\newcommand{\posterior}{\mathcal{P}}   
\newcommand{\likelihood}{\mathcal{L}}  
\newcommand{\prior}{\pi}               
\newcommand{\evidence}{\mathcal{Z}}    
\newcommand{\data}{D}        
\newcommand{\model}{\mathcal{M}}       
\newcommand{\params}{\theta}           
\newcommand{\KL}{\mathcal{D}_{\text{KL}}}    

\arxivnumber{2603.05472}
\title{\boldmath The Bayesian view of DESI DR2 with \texttt{unimpeded}: Evidence and tension in a combined analysis with CMB and supernovae across cosmological models}

\author[1,2,3]{Dily Duan Yi Ong\note{Corresponding author.}}
\author[1,3]{David Yallup}
\author[1,3]{and Will Handley}
\affiliation[1]{Kavli Institute for Cosmology, University of Cambridge,\\Madingley Road, Cambridge, CB3 0HA, U.K.}
\affiliation[2]{Cavendish Laboratory, University of Cambridge,\\J.J. Thomson Avenue, Cambridge, CB3 0HE, U.K.}
\affiliation[3]{Institute of Astronomy, University of Cambridge,\\Madingley Road, Cambridge, CB3 0HA, U.K.}

\emailAdd{dlo26@cam.ac.uk}

\abstract{We apply the \texttt{unimpeded} framework to perform a fully Bayesian reanalysis of the DESI DR2 data, using nested sampling with \texttt{PolyChord} to compute evidences for $\Lambda$CDM and seven extensions across combinations of DESI DR1/DR2, Planck CMB, supernovae (Pantheon+, Union3, DES-SN5YR, DES-Dovekie), and DES-Y1 weak lensing. The Bayesian Ockham's razor penalises extended models, yielding weaker or opposite preferences compared to $\Delta\chi^2$-based analyses. For DESI DR2 BAO combined with Planck CMB alone, the DESI collaboration's $3.1\sigma$ frequentist preference for $w_0w_a$CDM is eliminated entirely: we obtain ${\ln B = -0.57{\scriptstyle\pm0.26}}$, modestly favouring $\Lambda$CDM. Adding DES-Dovekie, the recalibration of DES-SN5YR, maintains this concordance (${\ln B = -0.30{\scriptstyle\pm0.19}}$). However, when the earlier DES-SN5YR calibration is included instead, the DESI collaboration's $4.2\sigma$ result survives the Bayesian Ockham penalty as a $3.07{\scriptstyle\pm0.10}\,\sigma$ preference (${\ln B = +3.32{\scriptstyle\pm0.27}}$). That this signal persists despite the Ockham penalty makes the role of tension quantification essential: our analysis traced the preference to the earlier DES-SN5YR calibration, which introduced a $2.95{\scriptstyle\pm 0.04}\,\sigma$ conflict with DESI DR2 within $\Lambda$CDM --- a tension that stands out from the grid --- reduced to $1.96{\scriptstyle\pm 0.04}\,\sigma$ with the DES-Dovekie recalibration. With DES-Dovekie, the Bayesian evidence for dynamical dark energy vanishes.}

\begin{document}
\maketitle
\flushbottom

\section{Introduction}
\label{sec:introduction}

The second data release (DR2) from the Dark Energy Spectroscopic Instrument (DESI) collaboration presents the most precise measurements of Baryon Acoustic Oscillations (BAO) to date \cite{desi2025}. The primary analysis of these data reports up to 4.2$\sigma$ preference for dynamical dark energy ($w_0w_a$CDM) over the standard flat $\Lambda$ Cold Dark Matter ($\Lambda$CDM) model, based on a frequentist likelihood-ratio test statistic, with the strongest preference arising from the combination of DESI DR2 BAO, Planck cosmic microwave background (CMB) measurements, and DES-SN5YR supernovae. This preference prompted further investigation into a possible deviation from the standard cosmological model~\cite{CosmoVerseNetwork:2025alb}, though independent analyses have identified inconsistencies between DESI data releases~\cite{Efstathiou2025BAO} and raised concerns about the DES-SN5YR supernova calibration~\cite{efstathiou_evolving_2025}. DES-Dovekie~\cite{Popovic2025DovelkieCalib,Popovic2025Dovekie} is a recalibration of DES-SN5YR, combining additional tertiary standard stars with a more flexible calibration model. As the precision of cosmological data improves, robust methods for model selection and tension quantification are needed not only for navigating competing theoretical models, but also for identifying systematic errors in datasets before they propagate into spurious claims of new physics.

In this companion paper to our previous work \cite{CompanionPaper}, we provide a complementary analysis of the DESI DR2 and DR1 data from a Bayesian perspective, focusing explicitly on model comparison and the quantification of tensions between datasets (see also~\cite{Hergt2026Consistency} for a related analysis using CosmoPower). Building upon the methodology established in \cite{UnimpededPaper,UnimpededSoftware}, we perform an analysis using full nested sampling runs with PolyChord \cite{Handley2015PolychordI,Handley2015PolychordII}. Full nested sampling, made possible by computational grants (DP192 and DP264) on the DiRAC High-Performance Computing facility, allows for the direct calculation of the Bayesian evidence, the quantity for model selection, enabling comparison of competing cosmological scenarios. We examine eight distinct cosmological models, utilizing two DESI datasets (\texttt{bao.desi\_2024\_bao\_all} and \texttt{bao.desi\_dr2}) in combination with a range of external probes, including supernovae catalogues (Pantheon+, Union3, DES-SN5YR, and DES-Dovekie), weak lensing data (DES Y1), and multiple \textit{Planck} CMB likelihood configurations. We treat DES-Dovekie as the recalibration of DES-SN5YR, and retain the earlier DES-SN5YR calibration to demonstrate that previously reported preferences for $w_0w_a$CDM~\cite{CompanionPaper} were driven by the earlier calibration, subsequently revised by DES-Dovekie.

This paper provides an alternative Bayesian framework that complements the primary DESI analysis. We present four key results: (1) the full Bayesian evidences for the base $\Lambda$CDM model and 7 extensions, allowing for a direct comparison of their relative plausibility; (2) the normalized model posterior probabilities, which rank the models given the data; (3) a systematic quantification of inter-dataset tension for 25 pairwise and triplet dataset combinations using multiple statistical metrics (including suspiciousness $S$, information ratio $Q$, evidence ratio $R$, and Bayesian model dimensionality); and (4) a systematic comparison across all models that examines the cosmological implications of the DESI DR2 dataset.

\section{Theory}
\label{sec:theory}

We briefly review the Bayesian framework for parameter estimation, model comparison, and tension quantification. This section provides a condensed summary of the methods detailed in our previous work~\cite{UnimpededPaper}, to which we refer the reader for further details.

\subsection{Bayesian Inference and Parameter Estimation}
\label{ssec:bayes_inference}
Within a given model $\model$, Bayesian inference updates the probability distribution of its parameters $\params$ in light of new data $\data$. The posterior probability distribution, $\posterior(\params) \equiv \Prob(\params|\data, \model)$, is given by Bayes' theorem~\cite{2008ConPh..49...71T}:
\begin{equation}
    \posterior(\params) = \frac{\likelihood(\params)\prior(\params)}{\evidence},
    \label{eq:bayes_theorem}
\end{equation}
where $\prior(\params) \equiv \Prob(\params|\model)$ is the prior probability distribution, $\likelihood(\params) \equiv \Prob(\data|\params, \model)$ is the likelihood of the data given the parameters, and $\evidence$ is the Bayesian evidence:
\begin{equation}
    \evidence = \int \likelihood(\params) \prior(\params) \,d\params.
    \label{eq:evidence}
\end{equation}
The Kullback-Leibler (KL) divergence~\cite{kullback1951information} measures the information gain from the prior to the posterior:
\begin{equation}
    \KL = \int \posterior(\params) \log\frac{\posterior(\params)}{\prior(\params)}\,d\params.
    \label{eq:kl_divergence}
\end{equation}
The evidence admits an exact information-theoretic decomposition~\cite{2021PhRvD.103l3511H}:
\begin{equation}
    \ln \evidence = \langle \ln \likelihood \rangle_\posterior - \KL,
    \label{eq:evidence_decomposition}
\end{equation}
where $\langle \ln \likelihood \rangle_\posterior = \int \posterior(\params) \ln \likelihood(\params)\,d\params$ is the posterior-averaged log-likelihood. This decomposition makes explicit the role of the evidence as a balance between goodness-of-fit $\langle \ln \likelihood \rangle_\posterior$ and an Ockham penalty $\KL$: a model is rewarded for fitting the data well, but penalised in proportion to the information gained from prior to posterior, i.e.\ the degree to which the data have constrained the parameters.

\subsection{Model Comparison}
\label{ssec:theory_model_comparison}
The posterior probability for a model $\model_i$, given data $\data$ and a set of competing models, is calculated using Bayes' theorem~\cite{2008ConPh..49...71T}. This relates the model posterior to its evidence $\evidence_i$ and prior probability $\prior_i \equiv \Prob(\model_i)$:
\begin{equation}
  \Prob(\model_i|\data) = \frac{\Prob(\data|\model_i) \Prob(\model_i)}{\Prob(\data)} = \frac{\evidence_i\prior_i}{\displaystyle\sum_j \evidence_j\prior_j}.
\label{eq:model_posterior}
\end{equation}
Assuming uniform prior probabilities for all models under consideration ($\prior_i = \mathrm{constant}$), the expression simplifies such that the model posterior is determined solely by the ratio of its evidence to the total evidence:
\begin{equation}
    \Prob(\model_i|\data) = \frac{\evidence_i}{\displaystyle\sum_j \evidence_j}.
\label{eq:model_prob}
\end{equation}
This formulation yields the normalised posterior probability for each model, providing a direct ranking amongst a set of competitors, which is an advantage over pairwise Bayes factor comparisons~\cite{2008ConPh..49...71T}.

While our primary metric for multi-model comparison throughout this work is the normalised posterior probability, $\Prob(\model_i|\data)$, we also compute the pairwise log Bayes factor, $\ln B$, to facilitate a direct comparison with results in the literature that employ this metric. The log Bayes factor is defined as the difference in log-evidences:
\begin{equation}
    \ln B = \ln \mathcal{Z}_{\text{model 1}} - \ln \mathcal{Z}_{\text{model 2}}.
\label{eq:bayes_factor}
\end{equation}
For the specific case of comparing $w_0w_a$CDM against $\Lambda$CDM, we compute:
\begin{equation}
    \ln B = \ln \mathcal{Z}_{w_0w_a\mathrm{CDM}} - \ln \mathcal{Z}_{\Lambda\mathrm{CDM}},
\label{eq:bayes_factor_w0wa}
\end{equation}
where $\mathcal{Z}$ is the Bayesian evidence obtained from nested sampling. A positive value of $\ln B$ indicates a preference for the $w_0w_a$CDM model, while a negative value indicates a preference for $\Lambda$CDM.

To further aid comparison with frequentist hypothesis tests that report significances in units of $\sigma$, we apply a two-step conversion procedure adapted from Trotta~\cite{2008ConPh..49...71T}. First, we employ the relationship established by Sellke et al.~\cite{sellke_calibration_2001}, which provides an upper bound on $B$ for a given $p$-value. Inverting this for a measured Bayes factor yields a lower bound on the equivalent $p$-value (see also~\cite{2021PhRvD.103l3511H}):
\begin{equation}
B \leq \bar{B} = -\frac{1}{e p \ln p} \quad \text{for } p \le e^{-1},
\label{eq:bayes_to_pvalue}
\end{equation}
where $\bar{B}$ represents the maximum Bayes factor consistent with a given $p$-value. Second, we map this $p$-value to an equivalent Gaussian significance using the standard inverse cumulative distribution function:
\begin{equation}
\sigma = \Phi^{-1}(1-p/2),
\label{eq:pvalue_to_sigma}
\end{equation}
where $\Phi^{-1}$ denotes the inverse of the standard normal cumulative distribution. This procedure yields conservative upper bounds on the significances derived from our Bayes factors, enabling direct numerical comparison with frequentist likelihood ratio tests.

We emphasise, however, that such conversions must be interpreted with care~\cite{berger_testing_1987,sellke_calibration_2001,kipping_exoplaneteers_2025}, as the two frameworks embody distinct philosophical approaches to statistical inference. The Bayesian evidence naturally incorporates Ockham's razor by penalising models for their prior volume~\cite{2021PhRvD.103l3511H}, whereas frequentist test statistics evaluate fit quality at a single point in parameter space. Nevertheless, the Bayesian interpretation as betting odds (asking whether one would bet on a given model at the implied odds) provides an intuitive measure of evidential strength. In well-behaved regimes, one expects broad agreement between properly calibrated Bayesian and frequentist model selection procedures regarding which model is preferred, even if the quantitative strength of preference differs.

\subsection{Tension Quantification}
\label{ssec:theory_tension_quantification}
To quantify the statistical consistency between datasets, we employ a suite of metrics (see~\cite{UnimpededPaper} for our implementation and application). The evidence ratio statistic, $R$~\cite{Marshall2006}, compares the joint evidence ($\evidence_{AB}$) from datasets $A$ and $B$ to the product of their individual evidences:
\begin{equation}
    R = \frac{\evidence_{AB}}{\evidence_A \evidence_B}.
    \label{eq:R_statistic_full}
\end{equation}
Drawing from the framework of~\cite{Handley2019}, the information ratio, $Q$, measures the change in information gain via the Kullback-Leibler divergences ($\KL$):
\begin{equation}
    Q = \KL^A + \KL^B - \KL^{AB}.
\label{eq:information_ratio}
\end{equation}
The suspiciousness, $S$, then quantifies statistical conflict between the datasets:
\begin{equation}
    \log S = \log R - Q.
\label{eq:suspiciousness_def}
\end{equation}
These statistics are calibrated using the Bayesian model dimensionality, $d$~\cite{Handley_dimensionality_2019}, which estimates the effective number of parameters constrained by the data and is calculated from the variance of the posterior-weighted log-likelihood:
\begin{equation}
    \frac{d}{2} = \langle(\log \likelihood)^2\rangle_{\posterior} - \langle\log\likelihood\rangle_{\posterior}^2.
\label{eq:bayesian_dimensionality}
\end{equation}
Finally, a $p$-value is calculated assuming $d-2\log S$ follows a $\chi^2_d$ distribution~\cite{Handley2019}, and this is converted to an equivalent Gaussian significance $\sigma$:
\begin{equation}
  p = \int_{d-2\log S}^{\infty} \chi_d^2(x)\,\mathrm{d}x,
\label{eq:tension_probability}
\end{equation}
\begin{equation}
    \sigma = \sqrt{2}\,\mathrm{Erfc}^{-1}(p).
\label{eq:sigma_conversion}
\end{equation}

\subsection{The Look Elsewhere Effect}
\label{ssec:look_elsewhere_effect}
The look-elsewhere effect (LEE) concerns the increased probability of chance discoveries when conducting numerous statistical tests. Our analysis spans $N=248$ distinct model and dataset configurations, necessitating a correction for this multiplicity. To account for this, we establish a global significance threshold rather than adjusting individual $p$-values. This threshold is defined as the significance level at which one false positive is expected across $N$ independent tests under the null hypothesis of no tension, calculated as:
\begin{equation}
    \sigma_{\text{threshold}} = \sqrt{2}\,\mathrm{Erfc}^{-1}\left(\frac{1}{N}\right).
\label{eq:sigma_threshold_corrected}
\end{equation}
For our $N=248$ tests, this yields $\sigma_{\text{threshold}} \approx 2.88$. We therefore consider any tension statistic exceeding this value to be significant, as it surpasses the level expected from random fluctuations across our investigation. For a more detailed discussion of this statistical treatment, we refer the reader to \cite{UnimpededPaper}.

\section{Methodology}
\label{sec:methodology}

\subsection{Cosmological Datasets}
\label{ssec:cosmological_datasets}

Our analysis employs a diverse set of cosmological datasets spanning multiple observational probes. \Cref{tab:datasets} summarises the datasets used in this work along with their corresponding likelihood components as implemented in \texttt{Cobaya}~\cite{Torrado2021Cobaya}. For cosmic microwave background (CMB) measurements, we utilise Planck 2018 data analysed with two independent high-$\ell$ likelihoods: the Plik likelihood~\cite{Planck2020likelihoods} and the CamSpec likelihood~\cite{CamSpec2021}. Each can be combined with Planck CMB lensing data~\cite{Planck2020lensing}, and we also include CMB lensing as a standalone dataset. For baryon acoustic oscillations (BAO), we analyse both DESI DR1~\cite{desi2024dr1} and DESI DR2~\cite{desi2025} datasets. Our Type Ia supernova (SN Ia) datasets comprise Pantheon+~\cite{Brout2022PantheonPlus}, Union3~\cite{Rubin2023Union3}, and the DES Y5 supernova sample in both its DES-Dovekie recalibration~\cite{Popovic2025DovelkieCalib,Popovic2025Dovekie} and the earlier DES-SN5YR calibration~\cite{DES2024SN5YR}, subsequently revised by DES-Dovekie. Finally, we incorporate weak gravitational lensing data from DES Y1~\cite{Abbott2018}. For detailed descriptions of these datasets and their implementations, we refer the reader to \cite{UnimpededPaper}. For the $w$CDM and $w_0w_a$CDM models, our nested sampling analysis adopts the same prior ranges $w_0 \in [-3, 1]$, $w_a \in [-3, 2]$ on the dark energy equation of state parameters as those used by the DESI Collaboration~\cite{desi2025}. 

\begin{table*}[t]
\centering
\begin{tabular}{p{7cm}p{8cm}}
\hline\hline
\textbf{Dataset} & \textbf{Likelihood} \\
\hline\hline
\multicolumn{2}{l}{\textbf{Cosmic Microwave Background}} \\
\hline
Planck~\cite{Planck2020likelihoods,PlanckClik} & \texttt{planck\_2018\_lowl.TT} \\
& \texttt{planck\_2018\_lowl.EE} \\
& \texttt{planck\_2018\_highl\_plik.TTTEEE} \\
& \texttt{planck\_2018\_highl\_plik.SZ} \\[0.3ex]
Planck with CMB lensing~\cite{Planck2020likelihoods,Planck2020lensing,PlanckClik} & \texttt{planck\_2018\_lowl.TT} \\
& \texttt{planck\_2018\_lowl.EE} \\
& \texttt{planck\_2018\_highl\_plik.TTTEEE} \\
& \texttt{planck\_2018\_highl\_plik.SZ} \\
& \texttt{planck\_2018\_lensing.clik} \\[0.3ex]
CamSpec~\cite{CamSpec2021} & \texttt{planck\_2018\_lowl.TT} \\
& \texttt{planck\_2018\_lowl.EE} \\
& \texttt{planck\_2018\_highl\_CamSpec2021.TTTEEE} \\[0.3ex]
CamSpec with CMB lensing~\cite{CamSpec2021,Planck2020lensing} & \texttt{planck\_2018\_lowl.TT} \\
& \texttt{planck\_2018\_lowl.EE} \\
& \texttt{planck\_2018\_highl\_CamSpec2021.TTTEEE} \\
& \texttt{planck\_2018\_lensing.clik} \\[0.3ex]
CMB Lensing~\cite{Planck2020lensing} & \texttt{planck\_2018\_lensing.clik} \\
\hline
\multicolumn{2}{l}{\textbf{Baryon Acoustic Oscillations}} \\
\hline
DESI DR1~\cite{desi2024dr1} & \texttt{bao.desi\_2024\_bao\_all} \\[0.3ex]
DESI DR2~\cite{desi2025} & \texttt{bao.desi\_dr2} \\
\hline
\multicolumn{2}{l}{\textbf{Type Ia Supernovae}} \\
\hline
Pantheon+~\cite{Brout2022PantheonPlus} & \texttt{sn.pantheonplus} \\[0.3ex]
Union3~\cite{Rubin2023Union3} & \texttt{sn.union3} \\[0.3ex]
DES-Dovekie~\cite{Popovic2025DovelkieCalib,Popovic2025Dovekie} & \texttt{sn.desdovekie} \\[0.3ex]
DES-SN5YR~\cite{DES2024SN5YR} & \texttt{sn.desy5} \\
\hline
\multicolumn{2}{l}{\textbf{Weak Lensing}} \\
\hline
DES Y1~\cite{Abbott2018} & \texttt{des\_y1.joint} \\
\hline\hline
\end{tabular}
\caption{Cosmological datasets and their corresponding likelihood components used in the analysis. Datasets are grouped by observational type with references to the actual data packages and implementation repositories used. Likelihood names correspond to those used by \texttt{Cobaya}.}
\label{tab:datasets}
\end{table*}

\subsection{Cosmological Models}
\label{ssec:cosmological_models}

This analysis examines the eight cosmological models forming the baseline of the Planck Legacy Archive~\cite{Planck2018params}, each probing different aspects of the standard cosmological paradigm. \Cref{tab:cosmological_models} details the parameters varied in each model along with their prior ranges. All models share the six baseline $\Lambda$CDM parameters $H_0$, $\tau_{\text{reio}}$, $\Omega_b h^2$, $\Omega_c h^2$, $\log(10^{10}A_s)$, $n_s$, with extensions adding one or two additional parameters to test specific physical hypotheses. For detailed descriptions of each cosmological model, we refer the reader to \cite{UnimpededPaper}. The nested sampling analyses were performed using \texttt{PolyChord}~\cite{Handley2015PolychordI,Handley2015PolychordII} via the \texttt{Cobaya} framework~\cite{Torrado2021Cobaya}.

\begin{table}
\centering
\begin{tabular}{p{2.2cm}p{1.8cm}p{2.2cm}p{6.8cm}}
\hline\hline
\textbf{Model} & \textbf{Parameter} & \textbf{Prior range} & \textbf{Definition} \\
\hline
$\Lambda$CDM & $H_0$ & [20, 100] & Hubble constant \\
 & $\tau_{\text{reio}}$ & [0.01, 0.8] & Optical depth to reionization \\
 & $\Omega_b h^2$ & [0.005, 0.1] & Baryon density parameter \\
 & $\Omega_c h^2$ & [0.001, 0.99] & Cold dark matter density parameter \\
 & $\log(10^{10}A_s)$ & [1.61, 3.91] & Amplitude of scalar perturbations \\
 & $n_s$ & [0.8, 1.2] & Scalar spectral index \\
\hline
$\Omega_k\Lambda$CDM & $\Omega_k$ & [-0.3, 0.3] & Curvature density parameter (varying curvature) \\[0.5ex]
$w$CDM & $w$ & [-3, -0.333] & Constant dark energy equation of state \\[0.5ex]
$w_0w_a$CDM & $w_0$ & [-3, 1] & Present-day dark energy equation of state \\
 & $w_a$ & [-3, 2] & Dark energy equation of state evolution (CPL parameterisation) \\[0.5ex]
$m_\nu\Lambda$CDM & $\Sigma m_\nu$ & [0.06, 2] & Sum of neutrino masses (eV) \\[0.5ex]
$A_L\Lambda$CDM & $A_L$ & [0, 10] & Lensing amplitude parameter \\[0.5ex]
$n_{\text{run}}\Lambda$CDM & $n_{\text{run}}$ & [-1, 1] & Running of spectral index ($dn_s/d\ln k$) \\[0.5ex]
$r\Lambda$CDM & $r$ & [0, 3] & Scalar-to-tensor ratio \\
\hline
\end{tabular}
\caption{Cosmological models and their parameter prior ranges. The standard six-parameter $\Lambda$CDM model serves as the baseline, with extensions adding one or two parameters to test specific physical hypotheses. All models share the six baseline $\Lambda$CDM parameters $H_0$, $\tau_{\text{reio}}$, $\Omega_b h^2$, $\Omega_c h^2$, $\log(10^{10}A_s)$, $n_s$. For the $w$CDM and $w_0w_a$CDM models, the prior ranges match those used by the DESI Collaboration~\cite{desi2025}.}
\label{tab:cosmological_models}
\end{table}

\subsection{Nested sampling chains availability}
\label{ssec:chains}

All models are implemented using the Cobaya framework~\cite{cobayaascl,Torrado2021Cobaya}, which interfaces with the CAMB Boltzmann code~\cite{Lewis:1999bs}. Nested sampling chains were generated using \texttt{PolyChord}~\cite{Handley2015PolychordI,Handley2015PolychordII} for the eight cosmological models and all dataset combinations detailed in this paper. The chains are permanently available on Zenodo and accessible via the open-source, pip-installable \texttt{unimpeded} Python package~\cite{UnimpededPaper,UnimpededSoftware}.

\section{Results}
\label{sec:results}

\subsection{Model Comparison}
\label{ssec:results_model_comparison}

\begin{figure}[p]
\vspace{-3cm}
\centering
    \includegraphics[width=\textwidth]{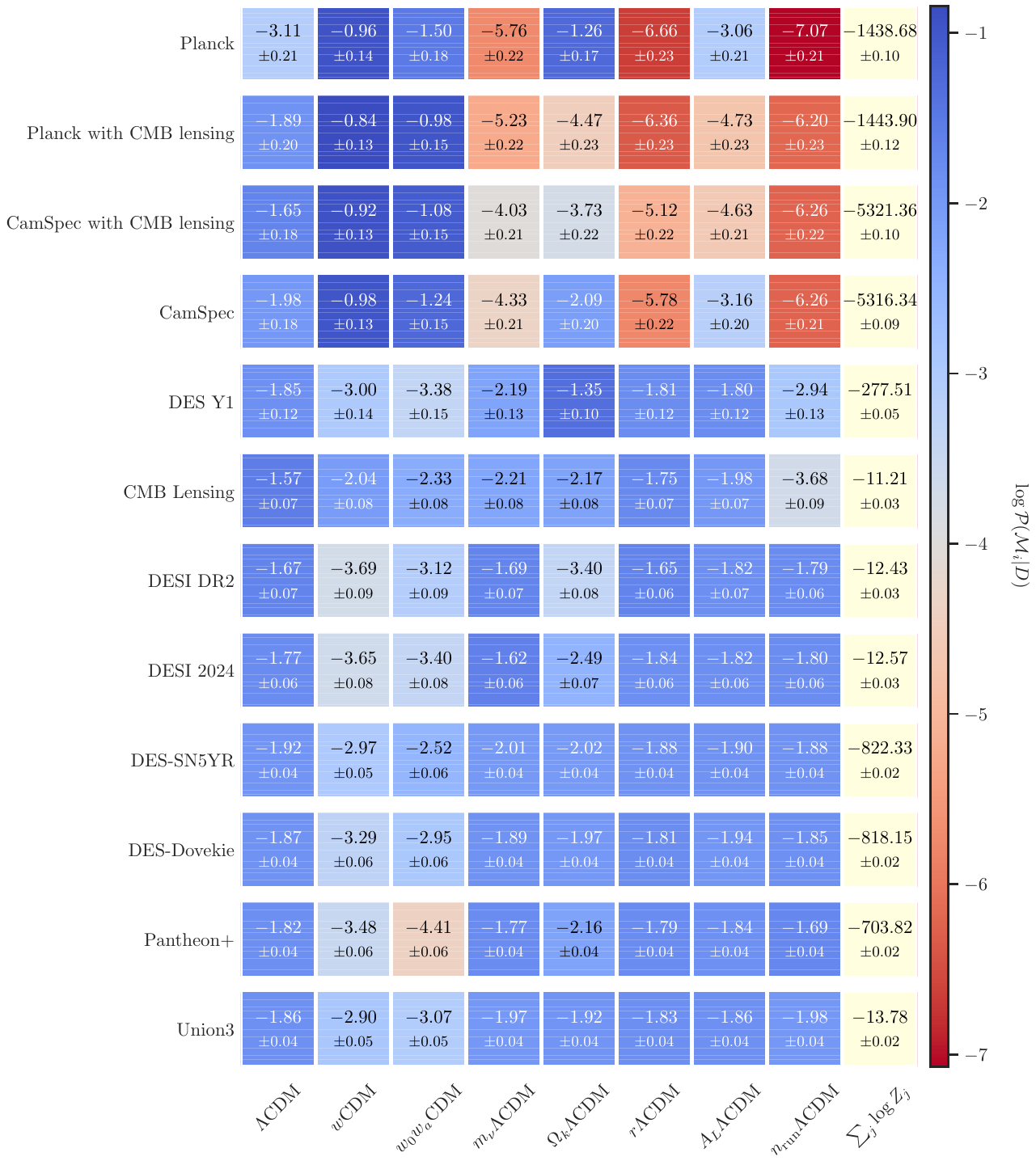}
    \caption{Log-posterior probabilities, $\log \Prob(\model_i|\data)$, for eight cosmological models tested against individual datasets. Higher probabilities (more evidence) are indicated by bluer shades. Models (columns) are arranged in ascending order by their constraining power ($\KL$ values) from Planck with CMB lensing, providing a consistent ordering across all model comparison figures. While various datasets show mild preferences for different model extensions, $\Lambda$CDM remains consistently well-supported. Model comparison is valid only along each row. The normalisation factor, $\log\left(\sum_j \evidence_j\right)$, is provided in the final column.}
    \label{fig:model_comp_single}
\end{figure}

\begin{figure}[p]
\vspace{-3cm}
\centering
    \includegraphics[width=0.95\textwidth]{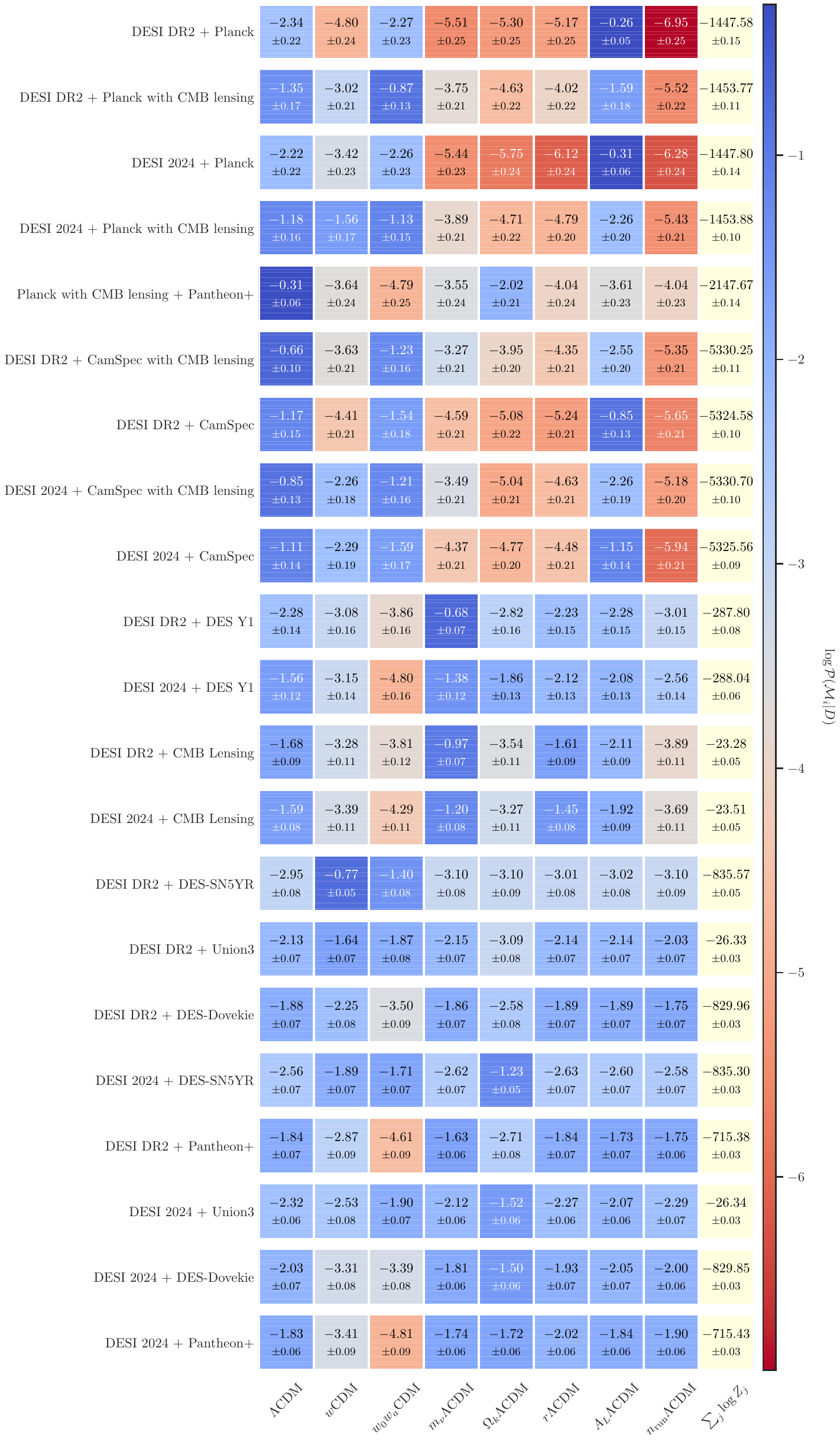}
    \caption{Model comparison results for paired dataset combinations, presented in the same format as \cref{fig:model_comp_single}.}
    \label{fig:model_comp_combo}
\end{figure}

\begin{figure}[p]
\vspace{-3cm}
\centering
    \includegraphics[width=\textwidth]{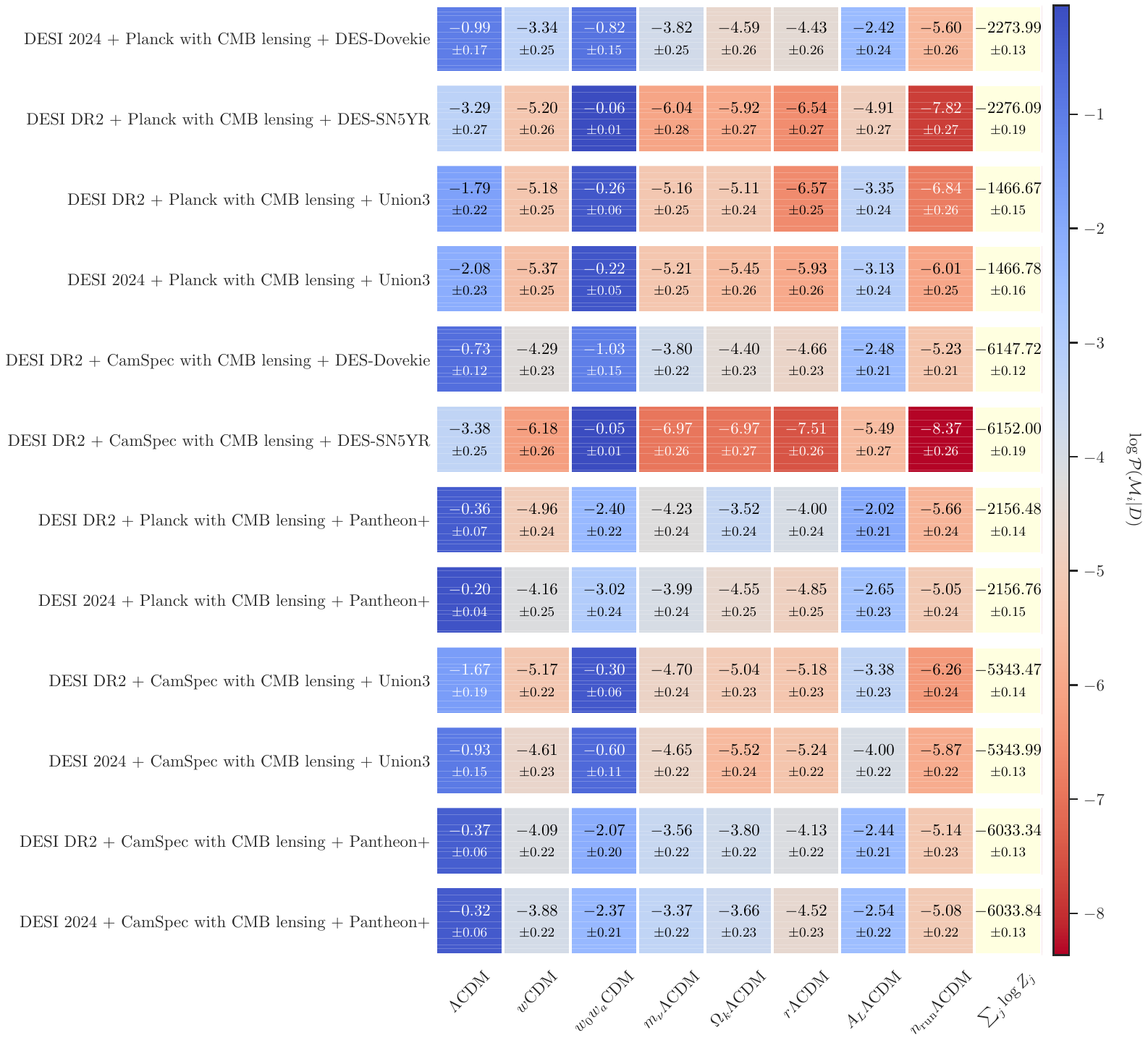}
    \caption{Model comparison results for triplet dataset combinations, following the format of \cref{fig:model_comp_single}. Triplet combinations generally support $\Lambda$CDM, with the exception of those involving DES-SN5YR (subsequently revised by the recalibrated DES-Dovekie), where the earlier calibration drives a preference for $w_0w_a$CDM.}
    \label{fig:model_comp_triplet}
\end{figure}

We perform a Bayesian model comparison to assess the relative performance of the eight cosmological models considered in this work (see \cref{ssec:theory_model_comparison} for the theoretical framework). For each model $\model_i$ for $i=0,1,...,N$ and dataset $\data$, we compute the normalised posterior probability $\Prob(\model_i|\data)=\evidence_i / \sum_j \evidence_j$ as defined in \cref{eq:model_prob}, because we adopted uniform prior $\prior = 1/N$. A higher (less negative) value indicates greater statistical support for a model, effectively rewarding its goodness-of-fit while penalising unnecessary complexity through the Ockham's razor principle inherent in the evidence calculation. For direct comparison with the literature that adopts a frequentist approach, we additionally compute the pairwise log Bayes factor and equivalent Gaussian significance $\sigma$ for the $w_0w_a$CDM versus $\Lambda$CDM comparison across all dataset combinations, following the conversion procedure outlined in \cref{ssec:theory_model_comparison}; these results are presented in \cref{tab:desi_comparison}. The full multi-model posterior probabilities for all eight models are displayed in \cref{fig:model_comp_single}--\ref{fig:model_comp_triplet}. The Bayesian evidence for any model can be recovered by multiplying the normalised posterior probability by the normalising factor $\log\left(\sum_j \evidence_j\right)$ displayed in the rightmost column of the corresponding row.

Our multi-model analysis reveals that model preference is sensitive to the combination of cosmological probes. As shown in \cref{fig:model_comp_single}, individual probes show distinct preferences: the DESI data (both DR1 and DR2) penalise the complexity of dynamical dark energy ($w_0w_a$CDM is disfavoured by $\Delta \ln P \approx -1.5$ relative to $\Lambda$CDM), while Planck CMB likelihoods favour it ($\Delta \ln P \approx +0.7$ to $+1.0$). The supernovae catalogues (Pantheon+, Union3, DES-Dovekie, DES-SN5YR) show little discriminatory power alone. When combined, the outcome depends on the choice of supernova dataset. Combinations involving Pantheon+ or the DES-Dovekie recalibration of DES-SN5YR consistently favour $\Lambda$CDM. For instance, for DESI DR2 + Pantheon+, $\Lambda$CDM is preferred, and for DESI DR2 + DES-Dovekie, $\Lambda$CDM remains preferred ($\ln P = -1.88{\scriptstyle\pm 0.07}$) over $w_0w_a$CDM ($\ln P = -3.51{\scriptstyle\pm 0.10}$). This preference persists in the three-probe combination: for DESI DR2 + CMB + DES-Dovekie, the pairwise Bayes factor is $\ln B =-0.30{\scriptstyle\pm 0.19}$, showing no evidence for $w_0w_a$CDM, and for DESI DR2 + CMB + Pantheon+, $\Lambda$CDM is favoured with $\ln P = -0.38{\scriptstyle\pm 0.06}$ versus $-2.06{\scriptstyle\pm 0.21}$ for $w_0w_a$CDM.

In contrast, a reversal occurs when using the earlier DES-SN5YR calibration or the Union3 catalogue. For the DESI DR2 + DES-SN5YR pair, preference shifts to $w$CDM ($\ln P = -0.77{\scriptstyle\pm 0.05}$) over $\Lambda$CDM ($\ln P = -2.96{\scriptstyle\pm 0.08}$). This effect is most pronounced in the three-probe analysis (\cref{fig:model_comp_triplet}). For the DESI DR2 + CMB + DES-SN5YR combination, $w_0w_a$CDM becomes the preferred model with $\ln P = -0.05{\scriptstyle\pm 0.01}$, while $\Lambda$CDM is disfavoured at $\ln P = -3.38{\scriptstyle\pm 0.25}$, a difference of $\Delta \ln P \approx +3.3{\scriptstyle\pm 0.25}$. The increased precision of DESI DR2 sharpens this dependence: for the Union3 triplet, the preference for $w_0w_a$CDM over $\Lambda$CDM widens from a marginal result with DR1 ($\ln P = -0.60{\scriptstyle\pm 0.11}$ vs. $-0.93{\scriptstyle\pm 0.15}$) to a preference with DR2 ($\ln P = -0.29{\scriptstyle\pm 0.06}$ vs. $-1.69{\scriptstyle\pm 0.20}$).

We now turn to a direct comparison of our Bayesian results with the frequentist analysis presented by the DESI collaboration~\cite{desi2025}. We compute the log Bayes factor (\cref{eq:bayes_factor_w0wa}) and its equivalent Gaussian significance for the same key dataset combinations as in Table VI of their work, following the methodology established in our companion letter~\cite{UnimpededPaper}. The results are presented in \cref{tab:desi_comparison}.

\begin{table*}[!htbp]
\small
\renewcommand{\arraystretch}{0.9}
\hspace{-1.5cm}
\begin{tabular}{@{}l@{\hspace{8pt}}r@{\hspace{6pt}}r@{\hspace{12pt}}r@{\hspace{6pt}}r@{}}
\toprule
& \multicolumn{2}{c@{\hspace{12pt}}}{This Work (Bayesian)} & \multicolumn{2}{c}{DESI Collab. (Frequentist)} \\
\cmidrule(lr){2-3} \cmidrule(l){4-5}
Dataset & $\ln B$ & Significance & $\Delta\chi^2_{\mathrm{MAP}}$ & Significance \\
\midrule
\multicolumn{5}{l}{\textbf{Individual Datasets}} \\
DESI DR2 & $-1.47{\scriptstyle\pm 0.11}$ & n/a & $-4.7$ & 1.7$\sigma$ \\
DESI DR1 & $-1.64{\scriptstyle\pm 0.10}$ & n/a & --- & --- \\
DES Y1 & $-1.55{\scriptstyle\pm 0.19}$ & n/a & --- & --- \\
CamSpec (lensing) & $+0.57{\scriptstyle\pm 0.26}$ & $1.67{\scriptstyle\pm 0.39}\,\sigma$ & --- & --- \\
CamSpec & $+0.74{\scriptstyle\pm 0.26}$ & $1.83{\scriptstyle\pm 0.21}\,\sigma$ & --- & --- \\
CMB Lensing & $-0.77{\scriptstyle\pm 0.11}$ & n/a & --- & --- \\
Planck (lensing) & $+0.89{\scriptstyle\pm 0.28}$ & $1.94{\scriptstyle\pm 0.19}\,\sigma$ & --- & --- \\
Planck & $+1.59{\scriptstyle\pm 0.27}$ & $2.34{\scriptstyle\pm 0.14}\,\sigma$ & --- & --- \\
DES-Dovekie & $-1.08{\scriptstyle\pm 0.08}$ & n/a & --- & --- \\
DES-SN5YR & $-0.60{\scriptstyle\pm 0.08}$ & n/a & --- & --- \\
Pantheon+ & $-2.59{\scriptstyle\pm 0.07}$ & n/a & --- & --- \\
Union3 & $-1.21{\scriptstyle\pm 0.07}$ & n/a & --- & --- \\
\midrule
\multicolumn{5}{l}{\textbf{Pairwise Combinations}} \\
DESI DR2 + CamSpec & $-0.38{\scriptstyle\pm 0.25}$ & n/a & $-9.7$ & 2.7$\sigma$ \\
DESI DR1 + CamSpec & $-0.50{\scriptstyle\pm 0.25}$ & n/a & --- & --- \\
DESI DR2 + CamSpec (lensing) & $-0.57{\scriptstyle\pm 0.26}$ & n/a & $-12.5$ & 3.1$\sigma$ \\
DESI DR1 + CamSpec (lensing) & $-0.38{\scriptstyle\pm 0.26}$ & n/a & --- & --- \\
DESI DR2 + Planck (lensing) & $+0.48{\scriptstyle\pm 0.27}$ & $1.54{\scriptstyle\pm 0.58}\,\sigma$ & --- & --- \\
DESI DR1 + Planck (lensing) & $+0.02{\scriptstyle\pm 0.26}$ & $0.06{\scriptstyle\pm 1.37}\,\sigma$ & --- & --- \\
DESI DR2 + Planck & $+0.07{\scriptstyle\pm 0.28}$ & $0.28{\scriptstyle\pm 1.36}\,\sigma$ & --- & --- \\
DESI DR1 + Planck & $-0.05{\scriptstyle\pm 0.27}$ & n/a & --- & --- \\
DESI DR2 + CMB Lensing & $-2.14{\scriptstyle\pm 0.15}$ & n/a & --- & --- \\
DESI DR1 + CMB Lensing & $-2.71{\scriptstyle\pm 0.14}$ & n/a & --- & --- \\
DESI DR2 + DES Y1 & $-1.57{\scriptstyle\pm 0.21}$ & n/a & --- & --- \\
DESI DR1 + DES Y1 & $-3.24{\scriptstyle\pm 0.20}$ & n/a & --- & --- \\
DESI DR2 + Pantheon+ & $-2.77{\scriptstyle\pm 0.12}$ & n/a & $-4.9$ & 1.7$\sigma$ \\
DESI DR1 + Pantheon+ & $-2.98{\scriptstyle\pm 0.11}$ & n/a & --- & --- \\
DESI DR2 + Union3 & $+0.25{\scriptstyle\pm 0.12}$ & $1.39{\scriptstyle\pm 0.31}\,\sigma$ & $-10.1$ & 2.7$\sigma$ \\
DESI DR1 + Union3 & $+0.42{\scriptstyle\pm 0.11}$ & $1.59{\scriptstyle\pm 0.10}\,\sigma$ & --- & --- \\
DESI DR2 + DES-Dovekie & $-1.63{\scriptstyle\pm 0.12}$ & n/a & --- & --- \\
DESI DR1 + DES-Dovekie & $-1.37{\scriptstyle\pm 0.12}$ & n/a & --- & --- \\
DESI DR2 + DES-SN5YR & $+1.56{\scriptstyle\pm 0.12}$ & $2.33{\scriptstyle\pm 0.06}\,\sigma$ & $-13.6$ & 3.3$\sigma$ \\
DESI DR1 + DES-SN5YR & $+0.84{\scriptstyle\pm 0.11}$ & $1.92{\scriptstyle\pm 0.07}\,\sigma$ & --- & --- \\
Planck (lensing) + Pantheon+ & $-4.50{\scriptstyle\pm 0.28}$ & n/a & --- & --- \\
\midrule
\multicolumn{5}{l}{\textbf{Triplet Combinations}} \\
DESI DR2 + CamSpec (lensing) + Pantheon+ & $-1.70{\scriptstyle\pm 0.26}$ & n/a & $-10.7$ & 2.8$\sigma$ \\
DESI DR2 + CamSpec (lensing) + Union3 & $+1.37{\scriptstyle\pm 0.27}$ & $2.23{\scriptstyle\pm 0.15}\,\sigma$ & $-17.4$ & 3.8$\sigma$ \\
DESI DR2 + CamSpec (lensing) + DES-Dovekie & $-0.30{\scriptstyle\pm 0.19}$ & n/a & --- & --- \\
DESI DR2 + CamSpec (lensing) + DES-SN5YR & $+3.32{\scriptstyle\pm 0.27}$ & $3.07{\scriptstyle\pm 0.10}\,\sigma$ & $-21.0$ & 4.2$\sigma$ \\
DESI DR2 + Planck (lensing) + Pantheon+ & $-2.02{\scriptstyle\pm 0.28}$ & n/a & --- & --- \\
DESI DR2 + Planck (lensing) + Union3 & $+1.52{\scriptstyle\pm 0.28}$ & $2.31{\scriptstyle\pm 0.14}\,\sigma$ & --- & --- \\
DESI DR2 + Planck (lensing) + DES-SN5YR & $+3.22{\scriptstyle\pm 0.28}$ & $3.03{\scriptstyle\pm 0.10}\,\sigma$ & --- & --- \\
DESI DR1 + CamSpec (lensing) + Pantheon+ & $-2.07{\scriptstyle\pm 0.29}$ & n/a & --- & --- \\
DESI DR1 + CamSpec (lensing) + Union3 & $+0.32{\scriptstyle\pm 0.26}$ & $1.23{\scriptstyle\pm 0.88}\,\sigma$ & --- & --- \\
DESI DR1 + Planck (lensing) + Pantheon+ & $-2.83{\scriptstyle\pm 0.33}$ & n/a & --- & --- \\
DESI DR1 + Planck (lensing) + Union3 & $+1.84{\scriptstyle\pm 0.32}$ & $2.46{\scriptstyle\pm 0.15}\,\sigma$ & --- & --- \\
DESI DR1 + Planck (lensing) + DES-Dovekie & $+0.17{\scriptstyle\pm 0.30}$ & $0.63{\scriptstyle\pm 1.29}\,\sigma$ & --- & --- \\
\bottomrule
\end{tabular}
\caption{Bayesian vs frequentist model comparison for $w_0w_a$CDM over $\Lambda$CDM. $\ln B$ is the log Bayes factor (positive favours $w_0w_a$CDM); Bayesian significance is computed only when $\ln B > 0$. DESI Collaboration $\Delta\chi^2_{\mathrm{MAP}}$ and frequentist significance are from Table~VI of Ref.~\cite{desi2025}. Combinations involving DES-SN5YR, subsequently revised by the recalibrated DES-Dovekie~\cite{Popovic2025DovelkieCalib,Popovic2025Dovekie}.}
\label{tab:desi_comparison}
\end{table*}

\begin{figure*}[p]
\vspace{-3cm}
\centering
    \includegraphics[width=\textwidth]{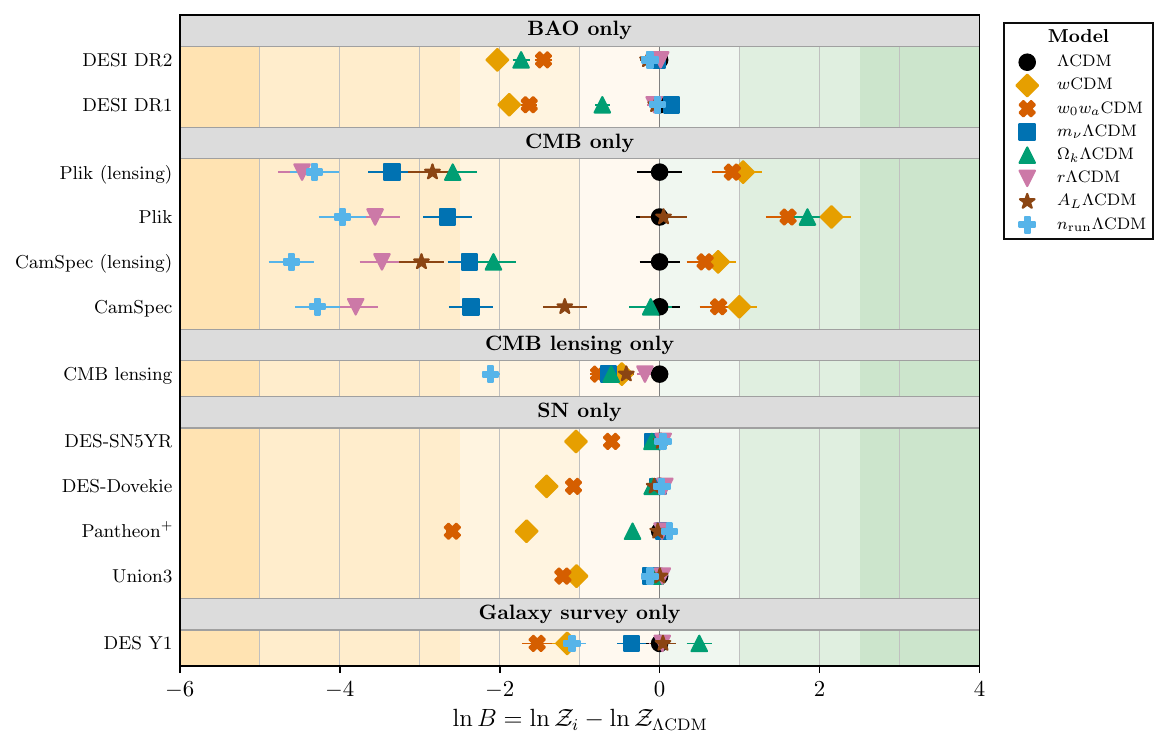}
    \caption{Visual summary of the Bayesian versus frequentist model comparison for the base model $\Lambda$CDM against 7 extension models using individual datasets, corresponding to \cref{tab:desi_comparison}.}
    \label{fig:bayes_factor_dot_single}
\end{figure*}

\begin{figure*}[p]
\vspace{-3cm}
\centering
    \includegraphics[width=\textwidth]{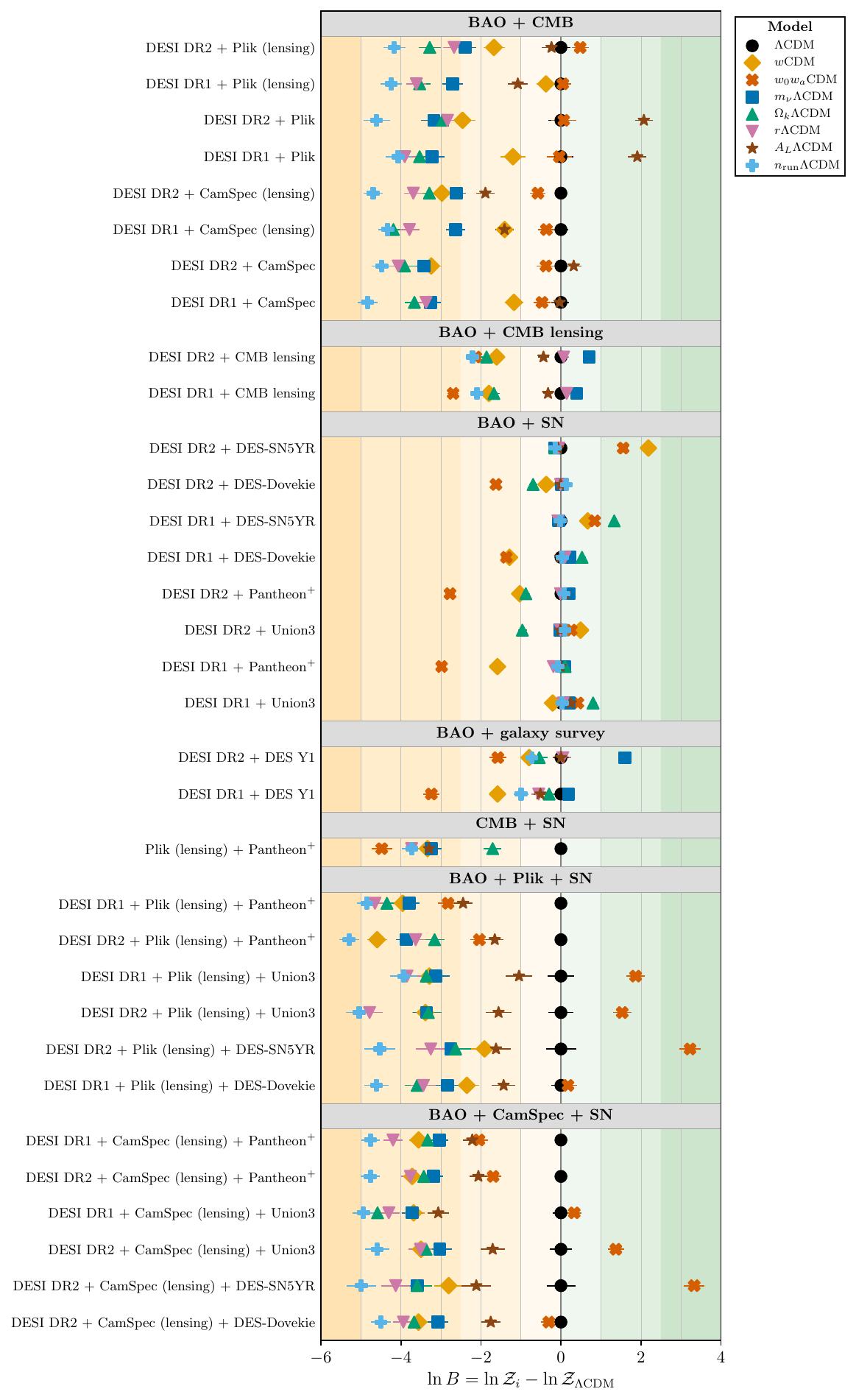}
    \caption{Visual summary of the Bayesian versus frequentist model comparison for the base model $\Lambda$CDM against 7 extension models using pairwise and triplet combinations, corresponding to \cref{tab:desi_comparison}.}
    \label{fig:bayes_factor_dot_combo}
\end{figure*}

Our Bayesian analysis consistently yields weaker or opposing conclusions compared to the frequentist $\Delta\chi^2_{\mathrm{MAP}}$ statistic, with the largest discrepancies arising for the combinations involving DES-SN5YR, subsequently revised by the recalibrated DES-Dovekie. For DESI data alone (DR1: $\ln B =-1.64$; DR2: $\ln B =-1.47$) or in pairwise combinations with the CMB (DESI DR2 + CamSpec: $\ln B =-0.57$), our results favour $\Lambda$CDM. In contrast, the DESI collaboration reports a preference for $w_0w_a$CDM for these same combinations (e.g., $\Delta\chi^2_{\mathrm{MAP}} = -4.7$ or 1.7$\sigma$ for DR2 alone, and $\Delta\chi^2_{\mathrm{MAP}} = -12.5$ or 3.1$\sigma$ for DESI DR2 + CamSpec). This opposing preference also holds for all combinations involving Pantheon+.

Agreement on the direction of preference for $w_0w_a$CDM is found only for combinations involving the Union3 or DES-SN5YR catalogues, though the strength of evidence differs substantially. For DESI DR2 + DES-SN5YR, our analysis gives $\ln B =+1.56$, while the frequentist preference is much stronger at $\Delta\chi^2_{\mathrm{MAP}} = -13.6$ (3.3$\sigma$). This supernova-driven preference culminates in the triplet combinations. For the DESI DR2 + CamSpec (lensing) + DES-SN5YR triplet, our analysis gives $\ln B =+3.32{\scriptstyle\pm 0.27}$ ($3.07{\scriptstyle\pm 0.10}\sigma$) --- a preference driven by DES-SN5YR (subsequently revised by the recalibrated DES-Dovekie) rather than genuine evidence for dynamical dark energy. This is also substantially weaker than the DESI collaboration's result of $\Delta\chi^2_{\mathrm{MAP}} = -21.0$ (4.2$\sigma$). Similarly, for the Union3 triplet, we find $\ln B =+1.37{\scriptstyle\pm 0.27}$ ($2.23{\scriptstyle\pm 0.15}\sigma$) compared to their $\Delta\chi^2_{\mathrm{MAP}} = -17.4$ (3.8$\sigma$).

Our finding that DESI DR2 combined with the CMB is consistent with $\Lambda$CDM is independently supported by the geometric analysis of Efstathiou~\cite{Efstathiou2025BAO}. The systematic discrepancy between our Bayesian and the DESI collaboration's frequentist results can be understood as an instance of the Jeffreys--Lindley paradox, which we explore in the following subsection.

\subsection{Relationship to frequentist significance}
\label{ssec:frequentist_significance}

The Jeffreys--Lindley paradox~\cite{jeffreys,lindley_statistical_1957} is a long-standing source of debate in the statistical literature, and it highlights a fundamental tension between Bayesian and frequentist approaches to hypothesis testing~\cite{wagenmakers_history_2022}. Unlike in the case of parameter estimation, where the received wisdom is that the two philosophical approaches should largely coincide given sufficiently informative data, when it comes to model comparison there is a known asymptotic discrepancy between the two paradigms. Numerous ``resolutions'' have been proposed to the paradox on both sides of the debate. The only clear consensus is that it is less a paradox and more a reflection of the different questions being asked by the two approaches~\cite{robert_jeffreys-lindley_2014}.

This paradox can be stated simply. When testing a point null hypothesis (e.g.\ $\Lambda$CDM) against a diffuse nested alternative (e.g.\ $w_0w_a$CDM), the observed frequentist $p$-value can become arbitrarily small, even for very small departures from the null, because the relevant test statistic typically grows with sample size. In contrast, the Bayesian evidence \emph{in favour} of the null hypothesis can be made arbitrarily large for fixed data by increasing the prior volume of the alternative model. A more diffuse prior dilutes the alternative's marginal likelihood via an Occam factor, even when the data strongly favour the alternative in terms of goodness-of-fit. This can lead to a situation where a frequentist test rejects the null hypothesis at a high significance level (e.g.\ $>2\sigma$), while the Bayesian evidence supports it (e.g.\ $\ln B < 0$, favouring the null under our sign conventions).

To establish that this is indeed the effect, we additionally explore the robustness of the frequentist significance quoted by the DESI collaboration. We take the simplest test case and investigate the DESI DR2 BAO-only dataset, which shows a $1.7\sigma$ significance to reject the null. Although the claims either way are not particularly strong for the DESI BAO data alone, we use this as a test case to probe these effects, and the computational effort is indicative of what is needed to robustly validate $>3\sigma$ claims.

\begin{figure}[t]
\centering
    \includegraphics{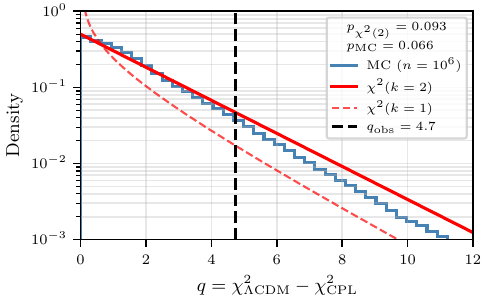}
    \caption{Monte Carlo validation of the frequentist test statistic $q = \chi^2_{\Lambda\mathrm{CDM}} - \chi^2_{\mathrm{CPL}}$ for DESI DR2 BAO alone. The empirical distribution under the $\Lambda$CDM null hypothesis (histogram) falls between the $\chi^2$ distributions for $k=1$ and $k=2$ degrees of freedom. Wilks' theorem overestimates the $p$-value: for $q_{\mathrm{obs}} \approx 4.7$, $p_{\mathrm{MC}} = 0.066$ versus $p_{\chi^2(2)} = 0.093$.}
    \label{fig:wilks_mc}
\end{figure}

We seek to confirm that the asymptotic formula for the $p$-value based on Wilks' theorem~\cite{wilks_large-sample_1938} is accurate by performing a Monte Carlo validation of the test-statistic distribution under the null hypothesis. Performing numerical validation of this kind is a standard that ought to be adhered to whenever significant results are claimed. Much of the attractive speed of the asymptotic formula is lost if one must perform a large number of simulations to validate it, and so it is often neglected in practice (the Bayesian analogue would be relying solely on a Laplace approximation to perform hypothesis testing). To perform this test one must first pick a fixed reference point for the nuisance parameters of the null (typically the maximum-likelihood estimates, here $\{H_0r_d, \Omega_M\}$). This reference point is then used to generate Monte Carlo \emph{toy datasets} from the Gaussian covariance of the DESI DR2 BAO likelihood. We generate $10^6$ toy datasets in this manner and fit each realisation to both $\Lambda$CDM and CPL ($w_0,w_a$) models via nonlinear least squares. We then compare the empirical distribution of $q \equiv -\Delta\chi^2_{\mathrm{MAP}}$ to the asymptotic $\chi^2(k{=}2)$ expectation, noting that for flat priors the MAP and ML are coincident. \Cref{fig:wilks_mc} compares this histogram to the $\chi^2$ distributions for $k=1$ and $k=2$ degrees of freedom, and displays the observed value $q_{\mathrm{obs}} \approx 4.7$ as a dashed vertical line. The empirical distribution falls between the $\chi^2$ distributions for $k=1$ and $k=2$ degrees of freedom, leading to a non-trivial adjustment of the significance of this test.

This numerical check confirms that the asymptotic $\chi^2(k{=}2)$ approximation is slightly conservative, with the Monte Carlo $p$-value being smaller than the Wilks estimate. This is extracted by integrating the tail probability of the test statistic exceeding $q_{\mathrm{obs}}$, leading to a Monte Carlo $p$-value of $p_{\mathrm{MC}} = 0.066$, in comparison to the Wilks estimate of $p_{\chi^2(2)} = 0.093$. This result is stable when varying the reference $\Lambda$CDM cosmology by $\pm 2\sigma$ along the degeneracy direction of the profile likelihood. We note that it would be prohibitive to perform this numerical check for more complex datasets due to the computational cost, primarily in evaluating the CMB likelihood.

Even in this simplest case we observe a divergence in conclusions. We find a $1.8\sigma$ (Monte Carlo) significance to reject the null, while the Bayes factor is $\ln B = -1.47$, favouring the null hypothesis. The evidence decomposition (\cref{eq:evidence_decomposition}) makes the source of this discrepancy explicit: the fit improvement ($\Delta\chi^2_{\mathrm{MAP}} \approx -4.7$) contributes $\approx 2.35$ to $\ln B$ in favour of $w_0w_a$CDM, but the $\KL$ penalty---quantifying how much the posterior contracts relative to the prior in the $(w_0, w_a)$ subspace---more than absorbs this, yielding a net preference for $\Lambda$CDM. For the DESI DR2 + CMB (lensing) combination, this effect is even more striking: $\Delta\chi^2_{\mathrm{MAP}} = -12.5$ ($3.1\sigma$ frequentist) is entirely absorbed by the Ockham penalty, producing $\ln B = -0.57$ (\cref{tab:desi_comparison}). Unlike in parameter estimation, where frequentist and Bayesian methods can provide reliable and consistent cross-checks~\cite{Herold:2025hkb}, model comparison results can be inconsistent, and one must be explicit about which inferential target is being addressed. The Jeffreys--Lindley paradox, which we believe explains this discrepancy, has existed as a philosophical question in statistical inference for over 70 years, and its persistence in the literature is a testament to the fact that there is no universally accepted resolution. Both sides invoke it as a criticism of the other, with refutations appearing on a regular basis~\cite{wagenmakers_history_2022}. We do not expect to resolve this debate here, but we highlight a few features that are relevant to the present discussion. First, it is important to emphasise that the frequentist hypothesis test makes no statement about the probability of the alternative model. It only quantifies the probability of observing data at least as extreme as the observed data under the null hypothesis. As such, claims of significant evidence \emph{for} evolving dark energy are not directly quantified by the frequentist analysis. Secondly, while the Bayesian evidence is sensitive to the choice of prior, in this case that sensitivity provides a useful safeguard. Resolutions to the paradox often focus on adapting significance thresholds~\cite{pericchia_adaptative_2016}, noting that as the number of observations grows, the significance threshold should in turn be adapted. In effect, this is already embodied in the particle physics literature as the famous $5\sigma$ threshold for discovery~\cite{Lyons:2013yja}, a post-hoc calibration to mitigate false discovery claims. We contend that the Bayesian framework provides a natural mechanism to require stronger data in order to claim a discovery in this nested scenario.

\subsection{Tension Quantification}
\label{ssec:tension_quantification_results}

\begin{figure}[p]
\vspace{-3cm}
\centering
    \includegraphics[width=0.95\textwidth]{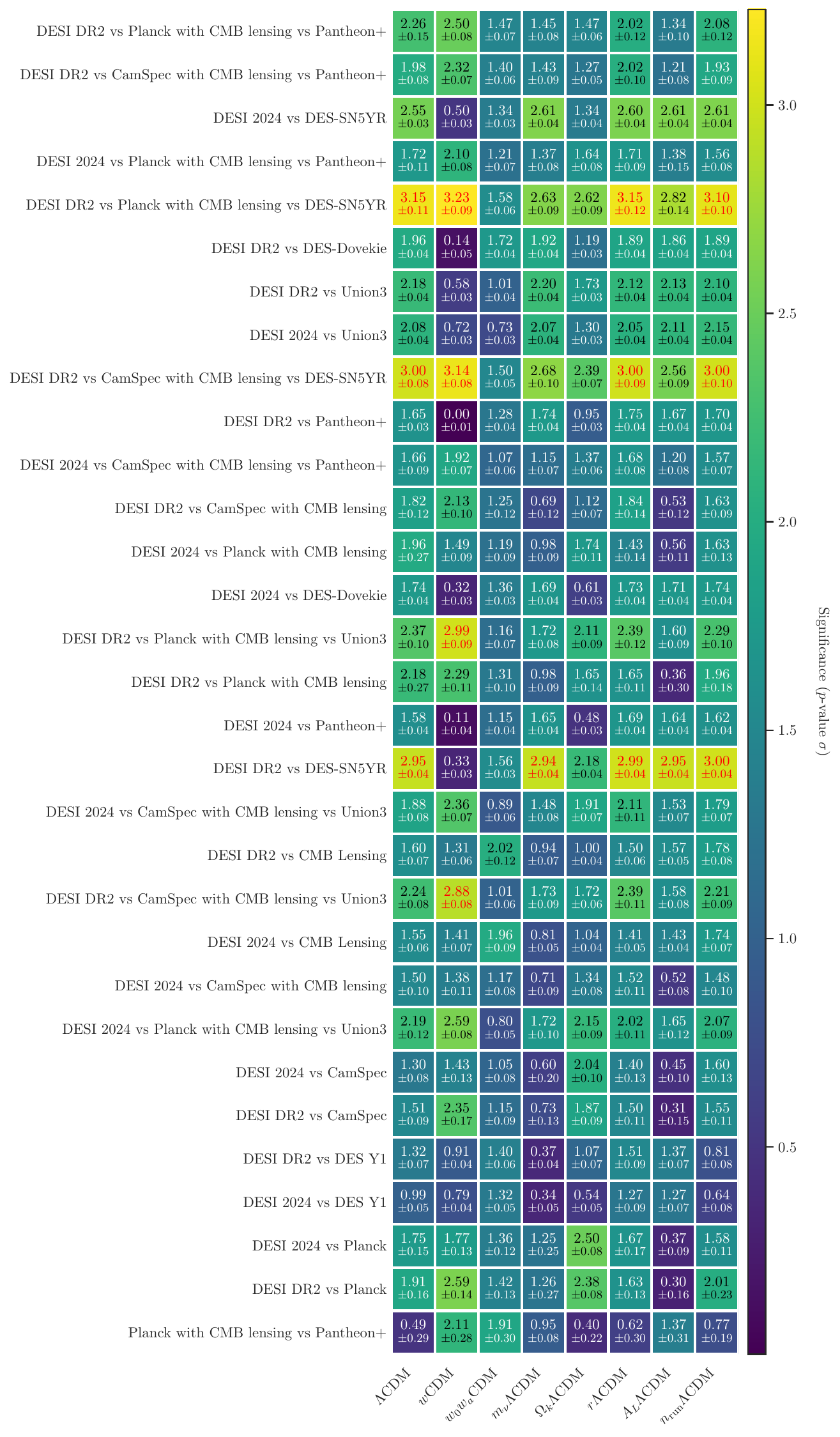}
    \caption{Tension significance ($\sigma$) for 25 dataset pairs across 8 models, sorted by descending average tension. Cells highlighted in red denote $\sigma > 2.88$, accounting for the look-elsewhere effect. Models (columns) are sorted by $\KL$ values from Planck with CMB lensing, consistent with all other figures.}
    \label{fig:tension_p}
\end{figure}

\begin{figure}[p]
\vspace{-3cm}
\centering
    \includegraphics[width=0.95\textwidth]{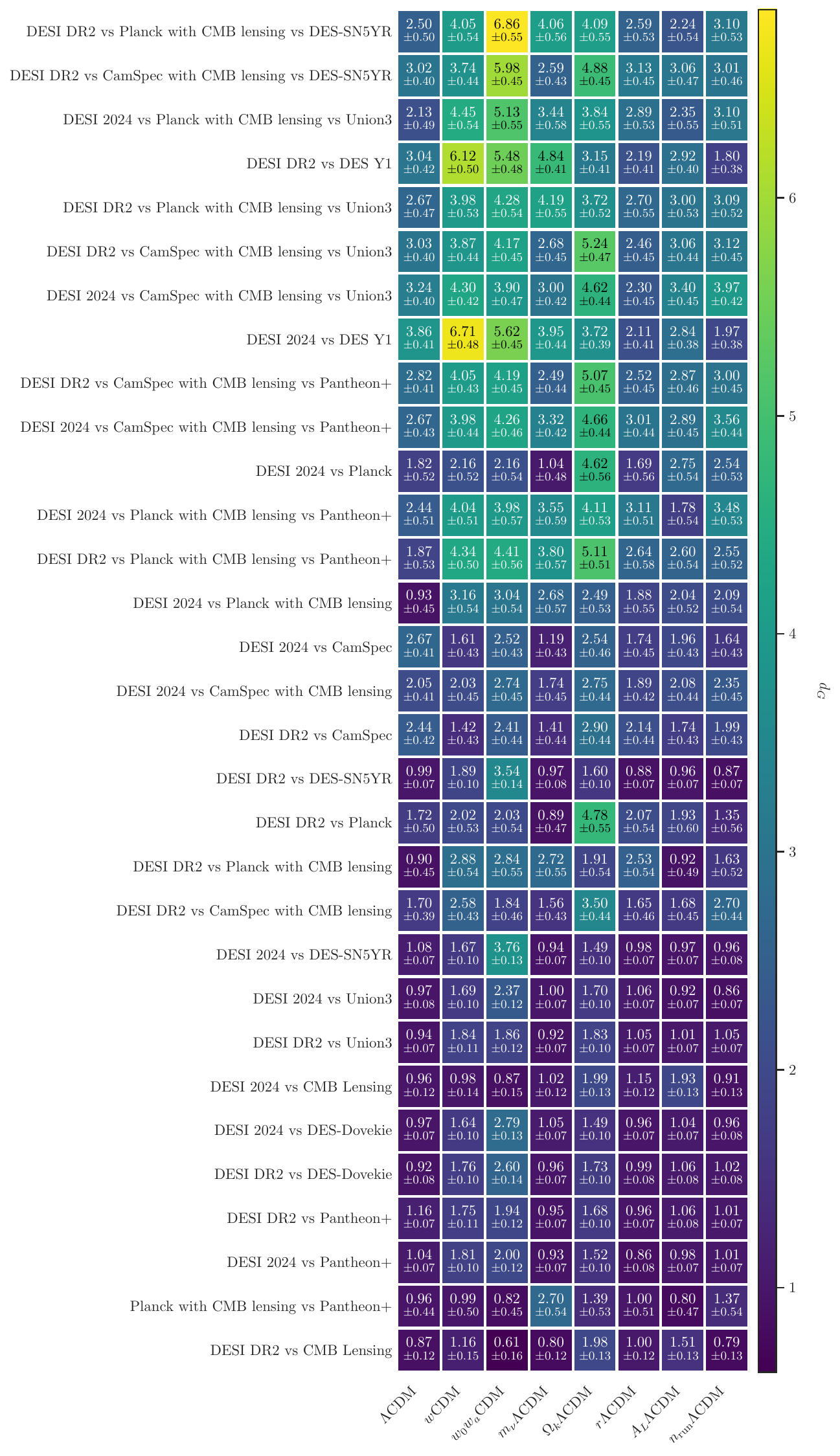}
    \caption{Bayesian Model Dimensionality ($d_G$) for all dataset pairs, sorted by ascending average dimensionality. Low values (top) indicate localised conflicts; high values (bottom) indicate broad, systemic disagreements.}
    \label{fig:tension_d_G}
\end{figure}

\begin{figure}[p]
\vspace{-3cm}
\centering
    \includegraphics[width=0.95\textwidth]{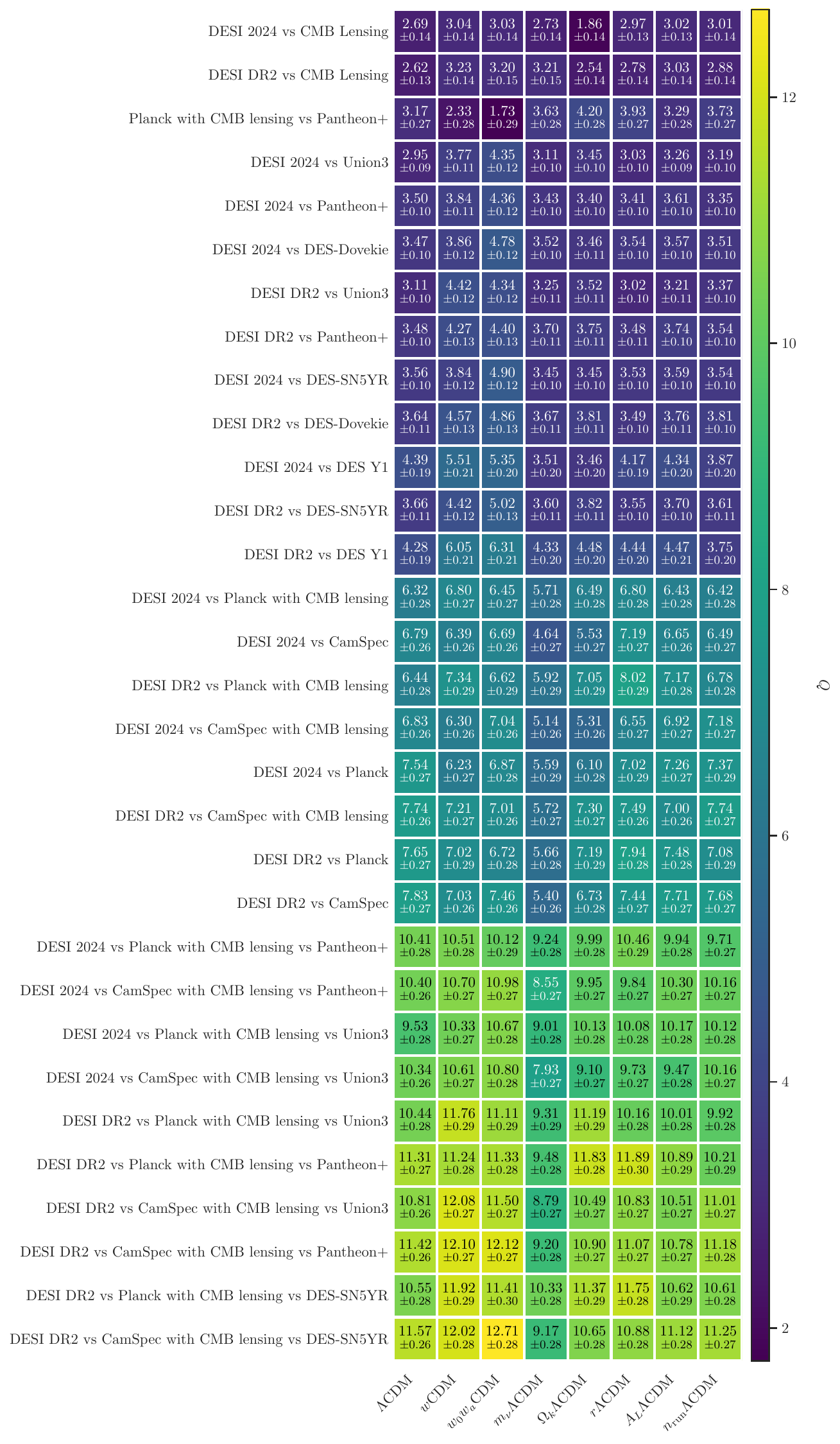}
    \caption{Information Ratio ($Q$) for all dataset pairs, sorted by ascending average value. Negative values indicate discrepant datasets whose combination provides less information gain than expected.}
    \label{fig:tension_I}
\end{figure}

\begin{figure}[p]
\vspace{-3cm}
\centering
    \includegraphics[width=0.95\textwidth]{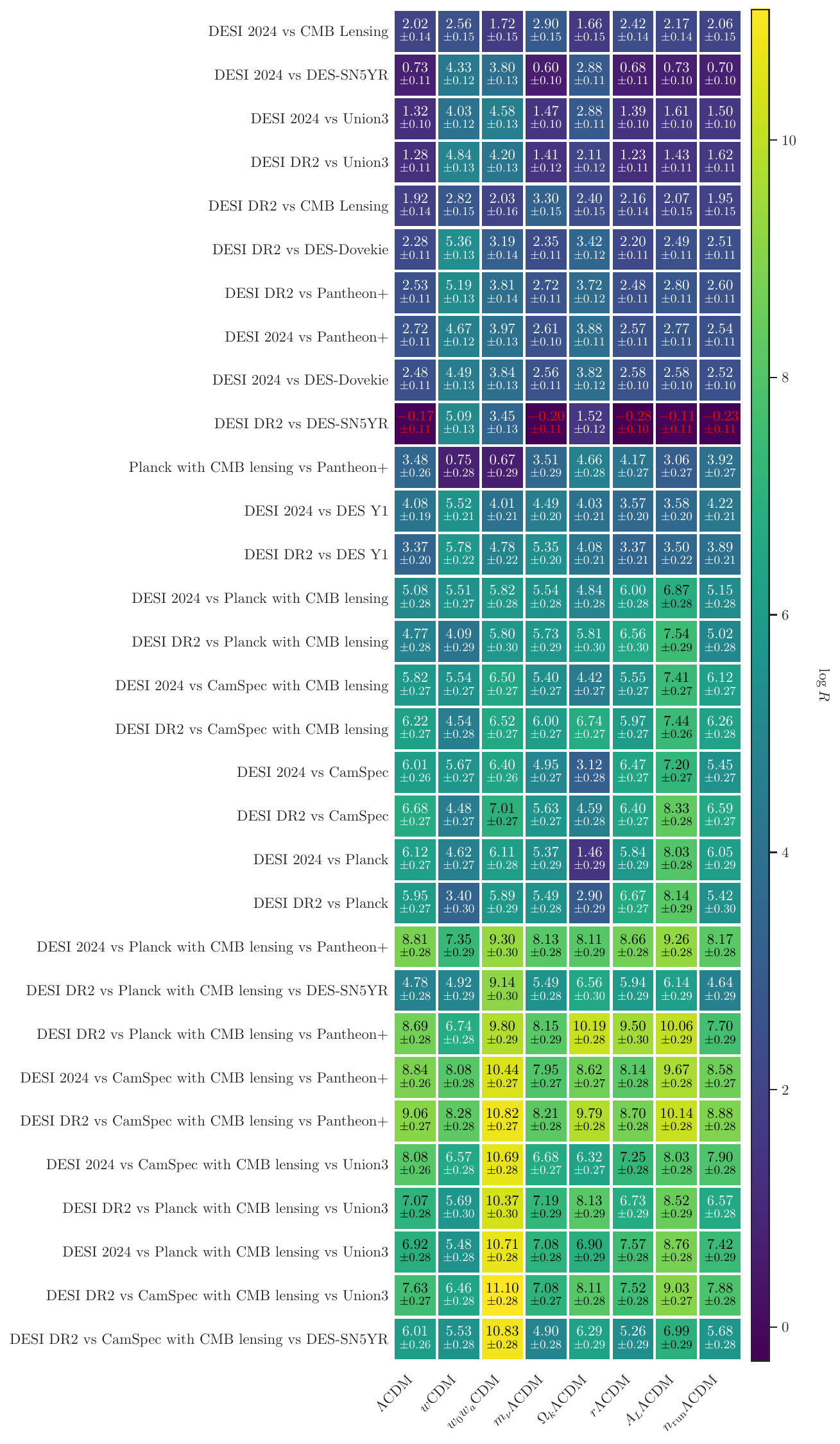}
    \caption{Logarithmic $R$ statistic ($\log R$) for all dataset pairs, sorted by ascending average value. Negative values (red) indicate suppressed joint evidence, signalling discordance.}
    \label{fig:tension_logR}
\end{figure}

\begin{figure}[p]
\vspace{-3cm}
\centering
    \includegraphics[width=0.95\textwidth]{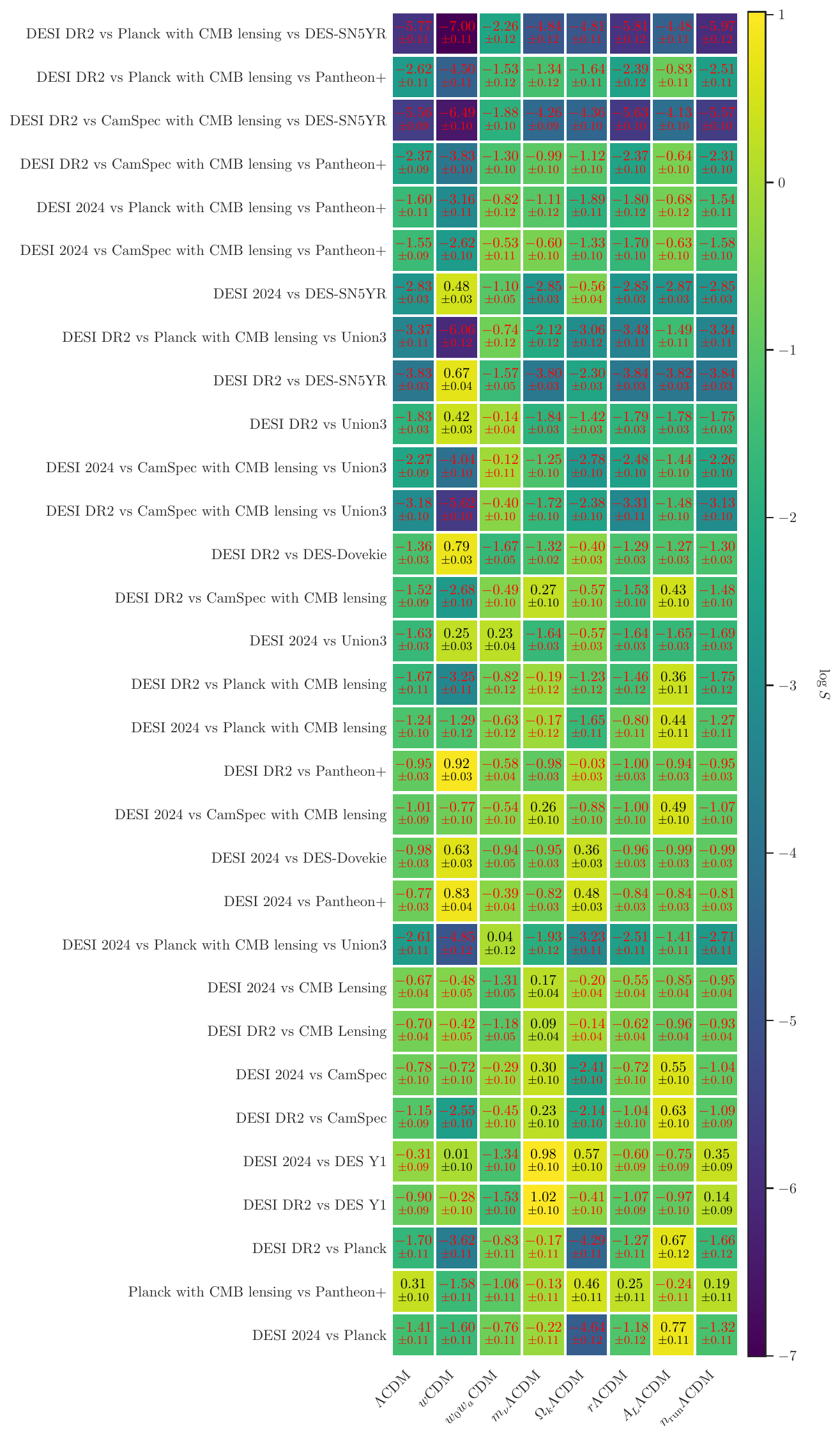}
    \caption{Logarithmic Suspiciousness ($\log S$) for all dataset pairs, sorted by ascending average value. Negative values (red) indicate tension, with more negative values corresponding to stronger conflicts.}
    \label{fig:tension_logS}
\end{figure}

We analyse the statistical consistency between pairs of cosmological datasets for each of the eight models under consideration. The analysis employs five distinct tension metrics computed using the \texttt{unimpeded} package~\cite{UnimpededPaper,UnimpededSoftware}, detailed in \cref{ssec:theory_tension_quantification}: the Gaussian significance ($\sigma$), the Bayesian model dimensionality ($d_G$), the information ratio ($Q$), the evidence ratio ($\log R$), and the suspiciousness ($\log S$). The results are summarised in a series of heatmaps (\Cref{fig:tension_p,fig:tension_d_G,fig:tension_I,fig:tension_logR,fig:tension_logS}). Following the statistical analysis of $N=248$ dataset and model combinations, we adopt a significance threshold of $\sigma > 2.88$ to account for the look-elsewhere effect, calculated by~\Cref{eq:sigma_threshold_corrected}. Tensions are also indicated by negative values for the $\log R$ and $\log S$ statistics. The Bayesian dimensionality, $d_G$, distinguishes between low-dimensional conflicts ($d_G \approx 1-2$), which are typically localised to a small parameter subspace, and more systemic, high-dimensional disagreements ($d_G > 3$).

The combination of DESI BAO data with Planck CMB measurements reveals generally mild tension, with significance values for $\Lambda$CDM ranging from $\sigma=1.50$ to $\sigma=2.18$. However, consistently negative suspiciousness values (e.g., $\log S = -2.68{\scriptstyle\pm 0.10}$ for \texttt{bao.desi\_dr2+planck\_2018\_CamSpec} in $w$CDM) flag a localised conflict at the likelihood level, providing an early diagnostic signal that warrants investigation into potential systematic origins. The low-to-moderate dimensionalities ($1.5 < d_G < 3.5$) indicate the disagreement spans a small number of parameter directions.

Using the recalibrated DES-Dovekie, the DESI DR2 BAO data show no significant tension with DES supernovae in $\Lambda$CDM: the \texttt{bao.desi\_dr2+sn.desdovekie} pair yields $\sigma = 1.96{\scriptstyle\pm 0.04}$, with $\log R = 2.28{\scriptstyle\pm 0.11}$ and $\log S = -1.36{\scriptstyle\pm 0.03}$. The mild residual disagreement is low-dimensional ($d_G = 0.92{\scriptstyle\pm 0.08}$). In contrast, when the earlier DES-SN5YR calibration~\cite{DES2024SN5YR} is used, a statistically significant tension emerges (\texttt{bao.desi\_dr2+sn.desy5}): $\sigma=2.95{\scriptstyle\pm 0.04}$ with $\log R = -0.17{\scriptstyle\pm 0.11}$ and $\log S \approx -3.8$, with a similarly low-dimensional structure ($d_G = 0.99{\scriptstyle\pm 0.07}$). This tension is absorbed by models with a dynamic dark energy equation of state, dropping to $\sigma=0.33{\scriptstyle\pm 0.03}$ in $w$CDM and $\sigma=1.56{\scriptstyle\pm 0.03}$ in $w_0w_a$CDM --- the mechanism by which the earlier DES-SN5YR calibration produced a spurious preference for dynamical dark energy.

The tension is sustained when Planck data are added to form the earlier DES-SN5YR triplet, \texttt{bao.desi\_dr2+planck\_2018\_CamSpec+sn.desy5}: ($\sigma \geq 3.00$ in four of the eight models) and its dimensionality increases ($d_G > 3$ for most models), indicating the calibration-driven conflict becomes more systemic. A separate comparison reveals that replacing DESI DR1 (\texttt{bao.desi\_2024\_bao\_all}) with the more precise DESI DR2 data systematically increases tension across all dataset combinations. For instance, in $\Lambda$CDM, the tension for the DR1 + Planck CamSpec + Union3 combination rises from $\sigma=1.88{\scriptstyle\pm 0.08}$ to $\sigma=2.24{\scriptstyle\pm 0.08}$ with DR2. While no DR1-based triplets surpass the $2.88\sigma$ threshold, the statistical power of DR2 pushes combinations involving the earlier DES-SN5YR calibration into the significant regime ($\sigma > 3.0$), amplifying the diagnostic signal from this calibration.

This analysis demonstrates that the Bayesian preference for dynamical dark energy, in the cases where such a preference arose, was driven by the extended model's capacity to absorb the tension introduced by the earlier DES-SN5YR calibration. The absence of this tension with the recalibrated DES-Dovekie confirms that the conflict was a calibration mismatch. For comparison, the DESI DR2 + Pantheon+ pair shows only mild tension in $\Lambda$CDM ($\sigma = 1.65{\scriptstyle\pm 0.03}$, $\log R = 2.53{\scriptstyle\pm 0.11}$) that is not significantly alleviated in extended models. The tension with DES-SN5YR, which was the largest among the supernova catalogues tested, is consistent with the recalibration of DES Y5 in~\cite{Popovic2025DovelkieCalib,Popovic2025Dovekie}, which combines additional tertiary standard stars with a more flexible calibration model.

\begin{figure*}[p]
\vspace{-3cm}
\centering
    \includegraphics[width=\textwidth]{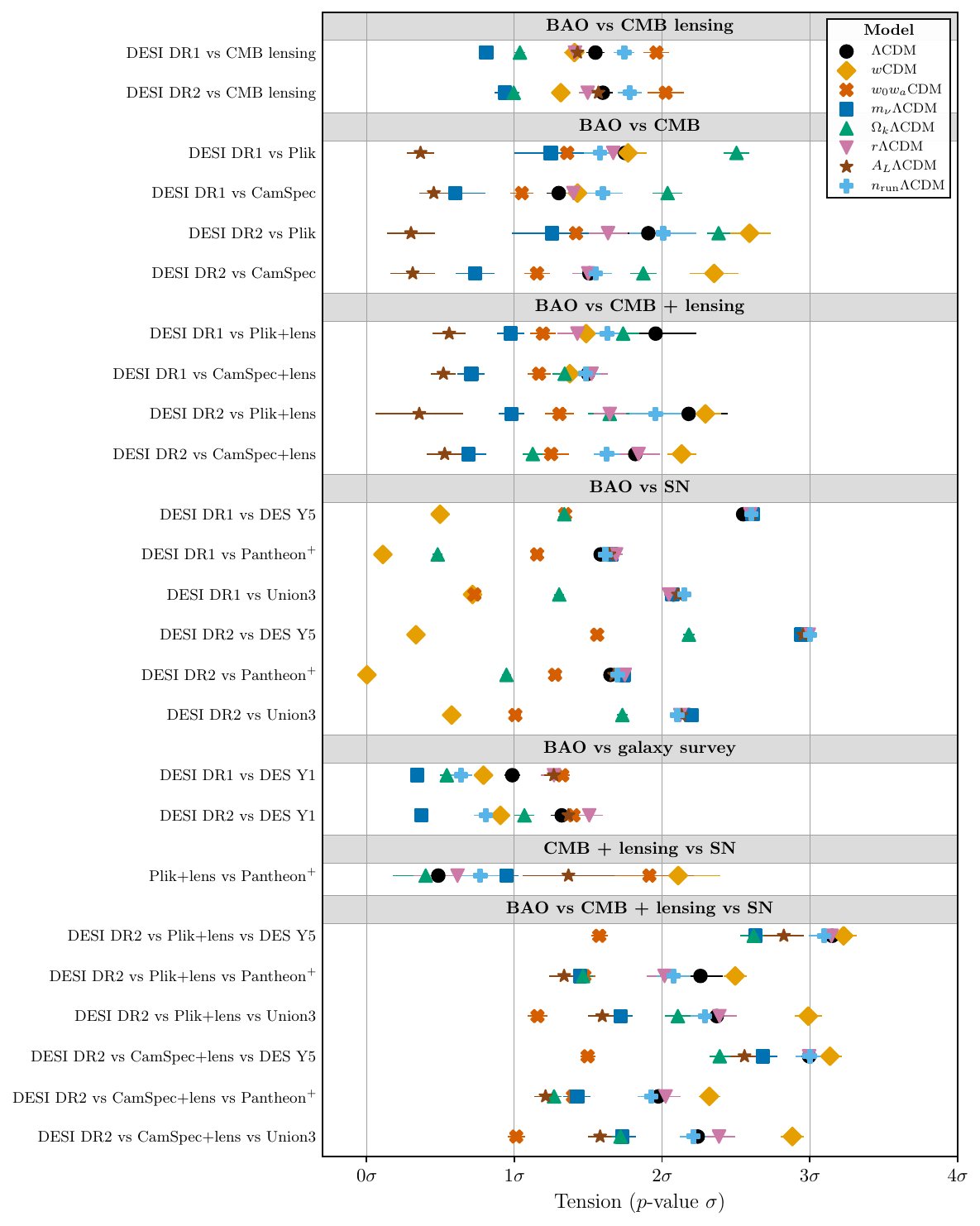}
    \caption{Visual summary of the tension significance ($\sigma$) for all dataset pairs across 8 cosmological models, corresponding to \cref{fig:tension_p}. Results are grouped by dataset pair category.}
    \label{fig:tension_dot}
\end{figure*}

\subsection{Constraining Power of DESI data on Models}
\label{ssec:constraining_power}

\begin{figure}[p]
\vspace{-3cm}
\centering
    \includegraphics[width=\textwidth]{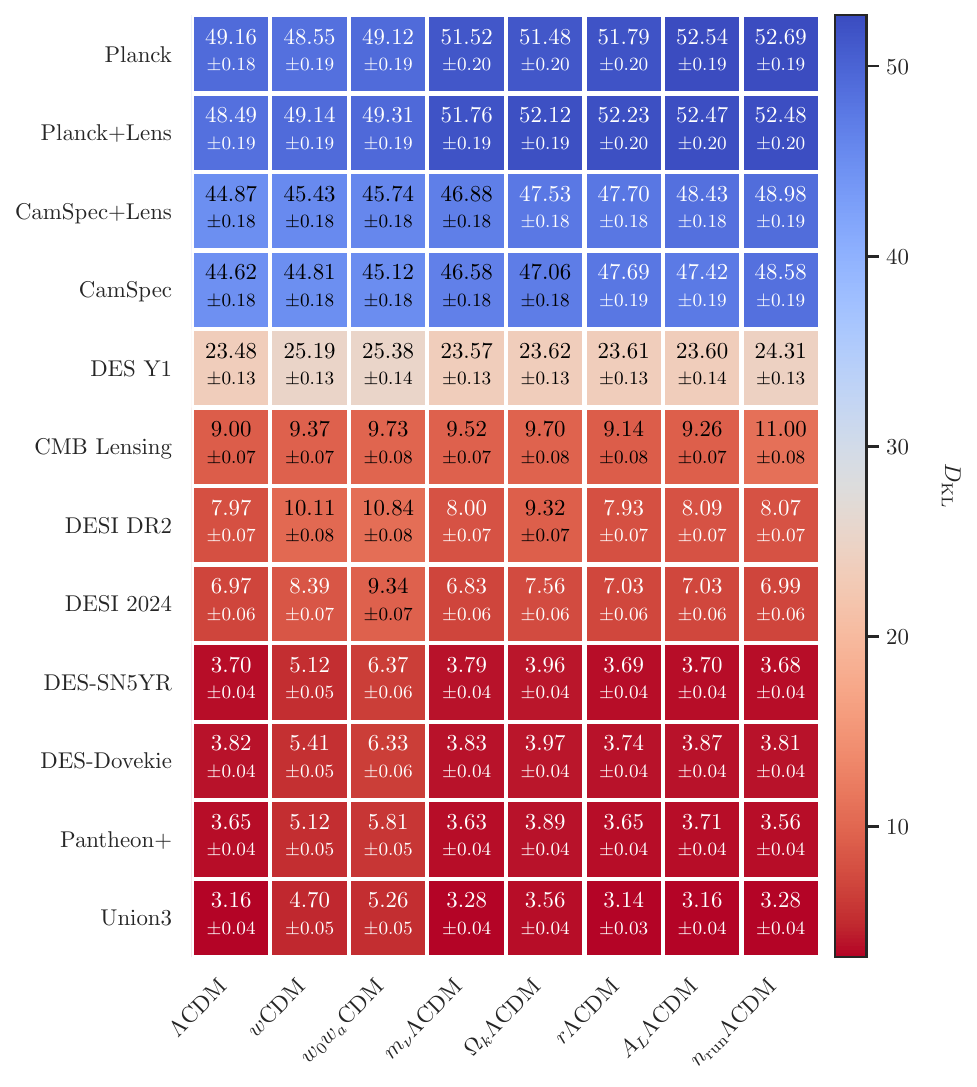}
    \caption{Heatmap of the Kullback-Leibler divergence ($\KL$), a metric for the constraining power of individual datasets. Datasets (rows) are ordered by their model-posterior-weighted average constraining power, $\langle \KL \rangle_{\Prob(\model)}$, while models (columns) are sorted in ascending order by their $\KL$ values from Planck with CMB lensing. The vertical gradient demonstrates that information gain is mainly determined by the dataset's statistical power rather than the specific cosmological model.}
    \label{fig:dkl_single}
\end{figure}

\begin{figure}[p]
\vspace{-3cm}
\centering
    \includegraphics[width=\textwidth]{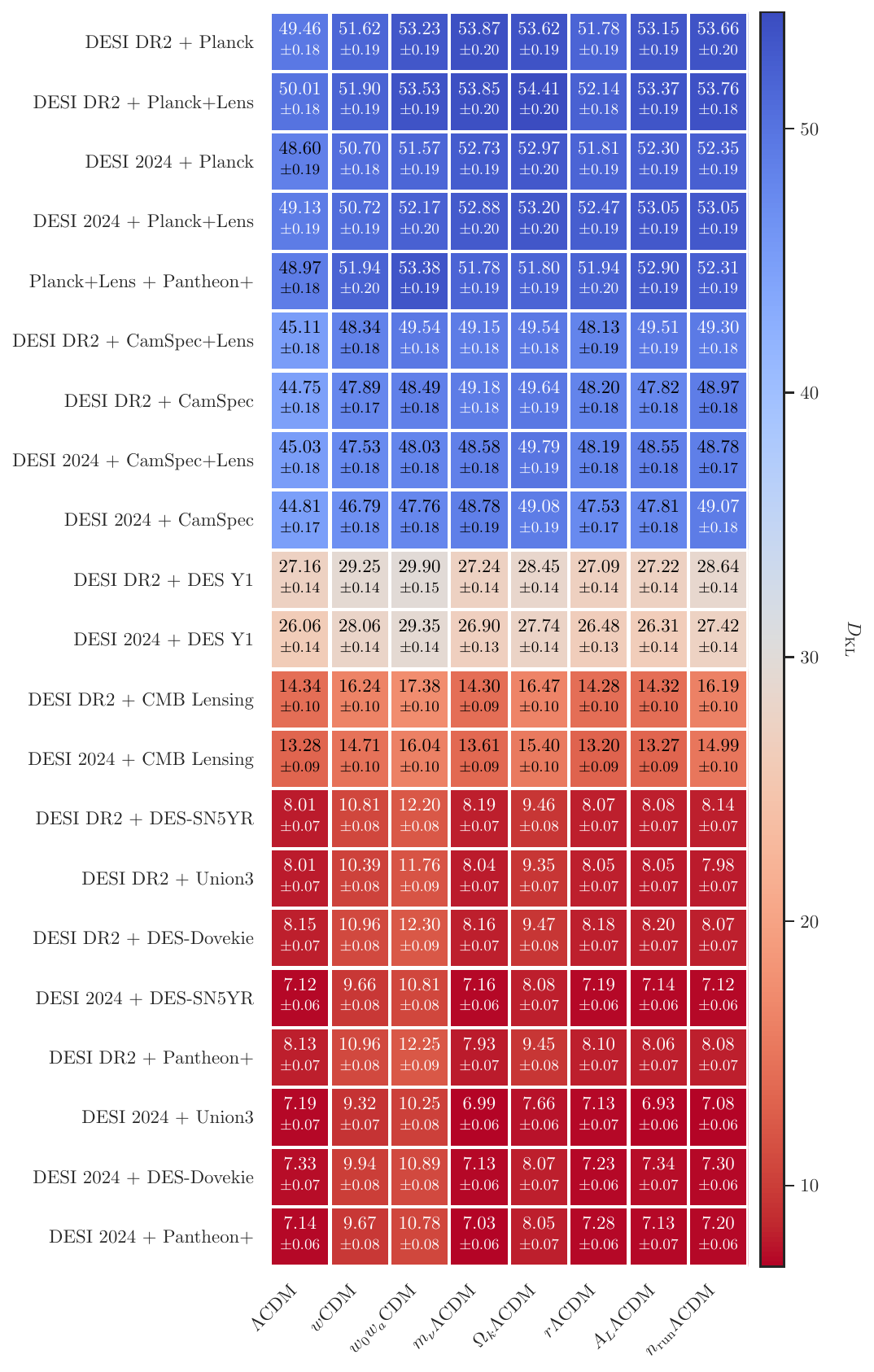}
    \caption{The Kullback-Leibler divergence ($\KL$) for paired dataset combinations. Combining datasets yields higher $\KL$ values than individual probes, reflecting increased constraining power. Models (columns) are sorted in ascending order by their $\KL$ values from Planck with CMB lensing, consistent with other figures.}
    \label{fig:dkl_combo}
\end{figure}

\begin{figure}[p]
\vspace{-3cm}
\centering
    \includegraphics[width=\textwidth]{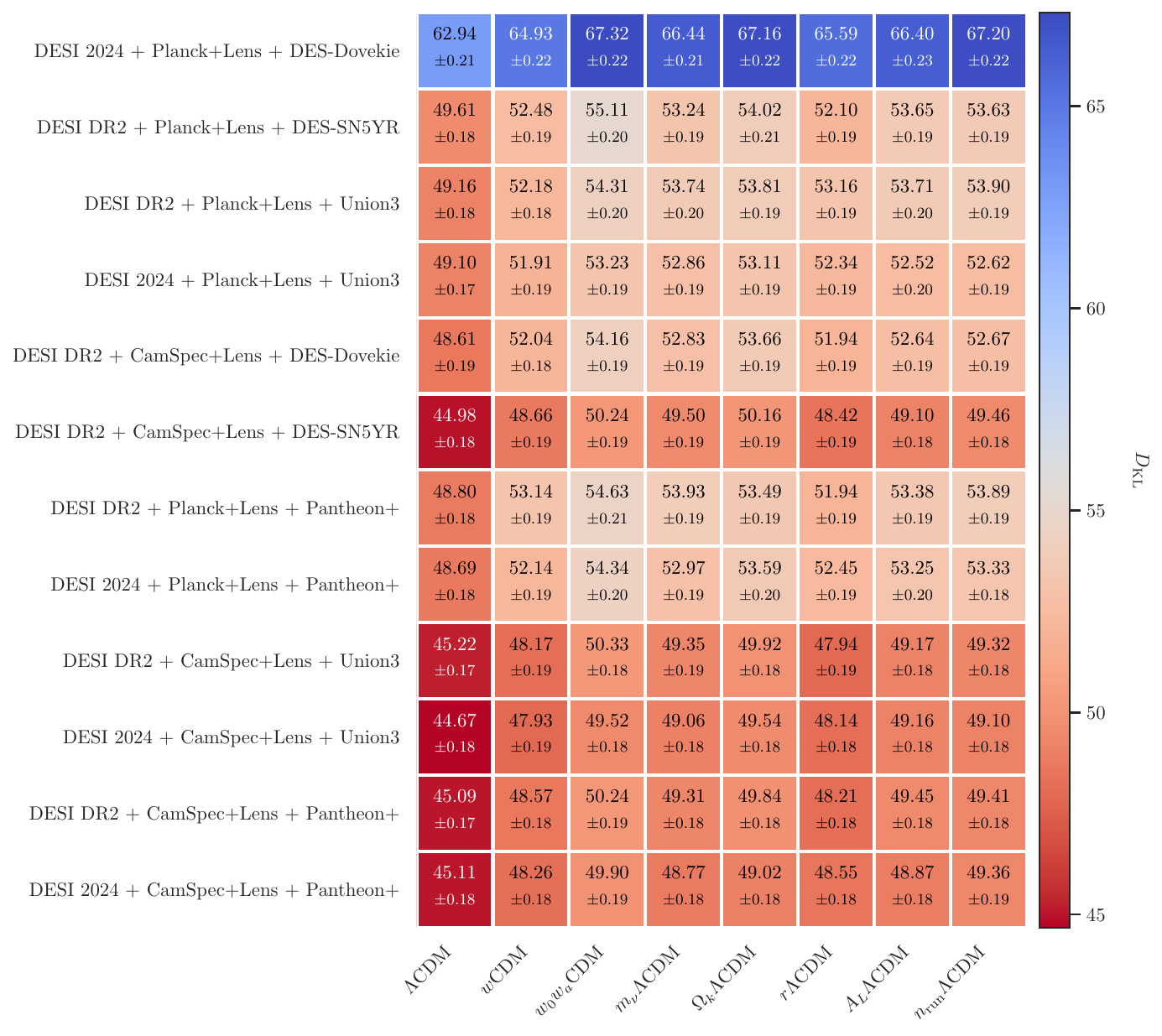}
    \caption{The Kullback-Leibler divergence ($\KL$) for triple dataset combinations. Combining three datasets further increases the $\KL$ values compared to single or paired combinations, demonstrating enhanced constraining power. The heatmap confirms that constraining power depends primarily on the dataset combination rather than the cosmological model. Models (columns) are sorted in ascending order by their $\KL$ values from Planck with CMB lensing, consistent with other figures.}
    \label{fig:dkl_triplet}
\end{figure}

We evaluate the statistical power of various datasets by calculating the Kullback-Leibler divergence ($\KL$), which measures the information gain from the prior to the posterior distribution. The results for individual, paired, and triple dataset combinations are presented as heatmaps in \Cref{fig:dkl_single,fig:dkl_combo,fig:dkl_triplet}, where larger $\KL$ values correspond to stronger parameter constraints.

To structure the visualisation, datasets (rows) are ranked by their overall constraining power, determined by the model-posterior-weighted average $\langle \KL \rangle_{\Prob(\model)}$. The models (columns) are sorted in ascending order based on their respective $\KL$ values from the Planck with CMB lensing dataset, which serves as a fixed reference for comparison across all figures.

\subsection{Comparison with concurrent analyses}
\label{ssec:hergt_comparison}

A concurrent and independent analysis by Hergt et al.~\cite{Hergt2026Consistency} performs Bayesian model comparison and tension quantification on overlapping data using CosmoPower emulators~\cite{Mancini2022CosmoPower} trained on \texttt{CLASS}, in contrast to our direct \texttt{CAMB} computation. Their analysis covers five models ($\Lambda$CDM, $\Omega_K$CDM, $w$CDM, $w_0w_a$CDM, $m_\nu\Lambda$CDM) across DESI DR2, multiple Planck CMB likelihoods (Plik, CamSpec, and Hillipop), and supernovae (Pantheon+, Union3, DES-SN5YR). The two analyses reach the same qualitative conclusions: the $w_0w_a$CDM preference is driven by the DES-SN5YR calibration, and vanishes with Pantheon+.

The analyses are complementary in several respects. Our $\ln B = -0.30{\scriptstyle\pm 0.19}$ for DESI DR2 + CMB + DES-Dovekie is broadly consistent with the Bayesian and frequentist results presented in~\cite{Popovic2025Dovekie}, providing an independent Bayesian cross-check using a different pipeline. This also complements the qualitative expectations of Hergt et al.~\cite{Hergt2026Consistency} for the recalibrated DES-Dovekie. Residual numerical differences with~\cite{Popovic2025Dovekie} are consistent with methodological choices: our priors are broader, they combine Planck+ACT+SPT, and differing CAMB settings can shift $\chi^2$ by $\sim 1$. We additionally track the DR1$\to$DR2 evolution, cover a broader model space (eight models including $A_L$, $n_{\mathrm{run}}$, and $r$ extensions), provide normalised multi-model posterior probabilities, and present the Jeffreys--Lindley / trials-factor analysis connecting Bayesian and frequentist results. Conversely, Hergt et al.~\cite{Hergt2026Consistency} include the Hillipop CMB likelihood and PR4 lensing not used here, provide historical context through SDSS DR12/DR16 BAO, and present detailed investigations of curvature tension, $\tau_{\mathrm{reio}}$ tension, and neutrino mass constraints.

Direct numerical comparison of Bayes factors between the two analyses requires care, as the prior ranges differ: our priors are broader (e.g.\ $w_0 \in [-3, 1]$ vs $[-2, 0]$; $H_0 \in [20, 100]$ vs $[40, 90]$~km\,s$^{-1}$\,Mpc$^{-1}$), leading to larger Ockham penalties. Nevertheless, that two independent pipelines --- \texttt{CAMB} vs \texttt{CLASS}/CosmoPower, with different prior choices --- reach the same conclusion strengthens the robustness of the result.

\clearpage

\section{Conclusions}
\label{sec:conclusions}

In this paper, we have presented a Bayesian analysis of the DESI DR2 dataset, providing a complementary perspective to the primary DESI collaboration results by focusing explicitly on model comparison and inter-dataset tension quantification. Utilising the \texttt{unimpeded} framework, we performed full nested sampling runs for eight distinct cosmological models across a range of single and combined datasets. This approach allowed us to compute the Bayesian evidence for each model-data combination, enabling an assessment of their relative plausibility that naturally incorporates the principle of Ockham's razor.

Our investigation yields several findings. First, for DESI DR2 combined with CMB data alone, the DESI collaboration's $3.1\sigma$ frequentist preference for $w_0w_a$CDM is eliminated by the Bayesian Ockham penalty: we find $\ln B =-0.57{\scriptstyle\pm 0.26}$, favouring $\Lambda$CDM. This is a direct consequence of the Jeffreys--Lindley paradox --- the look-elsewhere correction inherent to the Bayesian evidence absorbs the frequentist signal entirely. Second, using the recalibrated DES-Dovekie, we find that DESI DR2 and DES supernovae are consistent within $\Lambda$CDM ($\sigma=1.96{\scriptstyle\pm 0.04}$), and the three-probe combination DESI DR2 + CMB + DES-Dovekie yields $\ln B =-0.30{\scriptstyle\pm 0.19}$, showing no Bayesian evidence for $w_0w_a$CDM. Third, with DES-SN5YR (subsequently revised by the recalibrated DES-Dovekie), the DESI collaboration's $4.2\sigma$ result survives the Ockham penalty as $\ln B =+3.32{\scriptstyle\pm 0.27}$ ($3.07{\scriptstyle\pm 0.10}\,\sigma$). That this signal persists despite the Bayesian penalty is what makes the tension analysis essential: the tension metrics identified the source as a $2.95{\scriptstyle\pm 0.04}\,\sigma$ inter-dataset conflict introduced by the earlier DES-SN5YR calibration, rather than a physical signal. This diagnosis is reinforced by the subsequent DES-Dovekie recalibration of DES-SN5YR.

Our results demonstrate the value of Bayesian tension quantification as a diagnostic tool. The inter-dataset tension we identified pointed to DES-SN5YR as the source of the conflict, a diagnosis reinforced by the subsequent DES-Dovekie recalibration of DES-SN5YR. The previously reported preference for $w_0w_a$CDM was a consequence of the earlier DES-SN5YR calibration rather than a hint of new physics. All chains and analysis products from this work are publicly available via the \texttt{unimpeded} library. As future surveys deliver ever more precise data, this work demonstrates that Bayesian evidence and tension metrics provide a safeguard against mistaking dataset-level systematics for evidence of new physics.

\acknowledgments
The computations for this research were conducted on the Cambridge Service for Data Driven Discovery (CSD3). This work made use of the DiRAC component of CSD3, which is operated by the University of Cambridge Research Computing on behalf of the STFC DiRAC HPC Facility (www.dirac.ac.uk). Funding for the DiRAC facility is provided by BEIS capital funding via STFC capital grants ST/P002307/1 and ST/R002452/1, and by STFC operations grant ST/R00689X/1. DiRAC is a part of the National e-Infrastructure. W.H. is supported by a Royal Society University Research Fellowship.

\bibliographystyle{JHEP}
\bibliography{biblio}

\end{document}